\documentclass[11pt,a4paper,pdfa=true]{article}
\usepackage{jcappub}

\usepackage[T1]{fontenc}
\usepackage{graphicx,amssymb,amsmath,wasysym}

\newcommand{\hMsun}{{\ifmmode{h^{-1}{\rm
        {M_{\odot}}}}\else{$h^{-1}{\rm{M_{\odot}}}$}\fi}}  
\newcommand{\Msun}{{\ifmmode{{\rm
        {M_{\odot}}}}\else{${\rm{M_{\odot}}}$}\fi}} 
\newcommand{\rhosun}{{\ifmmode{
        {\rho_{\odot}}}\else{${\rho_{\odot}}$}\fi}}  
\newcommand{\Rsun}{{\ifmmode{{\rm
        {R_{\odot}}}}\else{${\rm{R_{\odot}}}$}\fi}}  
\newcommand{\hMpc}{{\ifmmode{h^{-1}{\rm Mpc}}\else{$h^{-1}$Mpc }\fi}}

\newcommand{\dd}{\text{d}}

\begin{document}

\hfill {\tt IFIC/14-35}

\title{Systematic uncertainties from halo asphericity in dark matter
  searches}

\author[1]{Nicol\'as Bernal,}
\emailAdd{nicolas@ift.unesp.br}

\author[2]{Jaime E. Forero-Romero,}
\emailAdd{je.forero@uniandes.edu.co}

\author[3]{Raghuveer Garani}
\emailAdd{garani@th.physik.uni-bonn.de}

\author[4]{and Sergio Palomares-Ruiz}
\emailAdd{sergio.palomares.ruiz@ific.uv.es}

\affiliation[1]{ICTP South American Institute for Fundamental
  Research, Instituto de F\'isica Te\'orica, Universidade Estadual
  Paulista, S\~ao Paulo, Brazil}
\affiliation[2]{Departamento de F\'{\i}sica, Universidad de los Andes,
  Cra. 1 No. 18A-10, Edificio Ip, Bogot\'a, Colombia}
\affiliation[3]{Bethe Center for Theoretical Physics and Physikalisches
  Institut, Universit\"at Bonn,\\Nu\ss allee~12, D-53115 Bonn,
  Germany}
\affiliation[4]{Instituto de F\'{\i}sica Corpuscular (IFIC), 
CSIC-Universitat de Val\`encia, Apartado de Correos 22085, E-46071
Val\`encia, Spain}

\abstract{
Although commonly assumed to be spherical, dark matter halos are 
predicted to be non-spherical by N-body simulations and their
asphericity has a potential impact on the systematic uncertainties in
dark matter searches.  The evaluation of these uncertainties is the
main aim of this work, where we study the impact of aspherical dark
matter density distributions in Milky-Way-like halos on direct and
indirect searches.  Using data from the large N-body cosmological
simulation {\texttt Bolshoi}, we perform a statistical analysis and
quantify the systematic uncertainties on the determination of local
dark matter density and the so-called $J$ factors for dark matter
annihilations and decays from the galactic center.  We find that, due
to our ignorance about the extent of the non-sphericity of the Milky
Way dark matter halo, systematic uncertainties can be as large as
35\%, within the 95\% most probable region, for a spherically averaged
value for the local density of 0.3-0.4~GeV/cm$^3$.  Similarly,
systematic uncertainties on the $J$ factors evaluated around the
galactic center can be as large as 10\% and 15\%, within the 95\% most
probable region, for dark matter annihilations and decays, respectively.
}

\maketitle

\section{Introduction}

Detecting and constraining the nature of dark matter (DM) is one of the
most pressing issues in physics today.  In contrast to collider
DM searches, direct and indirect detection methods crucially depend on
the properties of the DM halo of the Milky Way.  The number of nuclear
recoil events in direct detection experiments depends on the flux of
DM particles through the detector, which in turn depends on the local
DM density.  Searches for neutrinos from DM annihilations in the Sun
or Earth depend on the capture rate of DM particles, which also
depends on the flux of DM particles in the solar neighborhood, hence,
on the local DM density.  In addition, the flux of gamma-rays and
neutrinos produced in DM annihilations or decays in the galaxy depends
on the shape and orientation of the DM halo in the direction of
observation.

Most mass models of the Milky Way decompose the galaxy into three
components (the bulge, the disc and the dark halo), which are usually
described using analytical models constrained from observational data, 
such as terminal velocities in the inner galaxy, rotation velocities
in the outer galaxy, the solar position and velocity, the Oort's
constants, the mass at large radii, the local surface density,
etc.~\cite{Dehnen:1996fa, Klypin:2001xu, Catena:2009mf, Weber:2009pt,
  McMillan:2011wd, Nesti:2013uwa}.  The observational constraints for
the local DM density are of two types.  Some constraints are truly
local and are derived from stellar dynamics, whereas others are
constraints over an spherical average of the rotation curve.  These
two values are expected to be different in the case of a triaxial DM
halo, but most of the models of the galaxy have spherical symmetry.  

However, it is well known that N-body simulations predict
halos to be triaxial, being close to prolate in the inner part and
rounder at larger distances from the center~\cite{Frenk:1988zz,
  Katz:1991, Dubinski:1991bm, Warren:1992tr, Dubinski:1993df,
  Jing:2002np, Kasun:2004zb, Bailin:2004wu, Hopkins:2004np,
  Allgood:2005eu, Maccio':2006nu, Bett:2006zy, Hayashi:2006es,
  Kuhlen:2007ku, Stadel:2008pn, MunozCuartas:2010ig, VeraCiro:2011nb,
  Schneider:2011ed, Vera-Ciro:2014ita}, although with the
incorporation of baryons they predict halos to become
rounder~\cite{Kazantzidis:2004vu, Bailin:2005xq, Gustafsson:2006gr,
  Kazantzidis:2010jp, Bryan:2012mw}.  Observationally, early dynamical
studies using stellar kinematics and the variation in thickness of the
Milky Way's \texttt{HI} layer with radius, concluded that the DM halo
around the galactic disk is oblate~\cite{Olling:1999ss}.  On the other
hand, in the Milky Way, the  gravitational potential can be
constrained by using tidal streams of stars located at large distances
from the galactic center, where the potential is dominated by the DM
component.  This latter approach has been followed in recent years by
using the Sagittarius dwarf tidal stream~\cite{Ibata:2000pu,
  Helmi:2004id, Johnston:2004en, Law:2004ep, Fellhauer:2006mt,
  MartinezDelgado:2006bq, Law:2009yq, Law:2010pe, Deg:2012eu,
  Ibata:2012br, Vera-Ciro:2013nna}.  However, very different
conclusions have been reached depending on the adopted criterion.  An 
almost spherical halo was found in early works~\cite{Ibata:2000pu,
  Fellhauer:2006mt}, whereas an oblate halo has also been
claimed~\cite{Johnston:2004en, MartinezDelgado:2006bq, Law:2009yq,
  Law:2010pe, Deg:2012eu, Vera-Ciro:2013nna}, and even a prolate halo
has been suggested~\cite{Helmi:2004id}.  The triaxiality of the halo
was required in order to simultaneously fit the density profile and the
kinematics of the stream~\cite{Law:2009yq, Law:2010pe, Deg:2012eu}.
However, this almost oblate ellipsoid was found to have the minor axis
contained within the galactic disc, which represents an unstable
configuration~\cite{Debattista:2013rm}.  In addition, it has been
recently suggested that spherical solutions should be also allowed by
observational data~\cite{Ibata:2012br}.  Thus, the debate about the
shape and configuration of the DM halo of the Milky Way is open and a 
final consensus is far from being reached.  Therefore, in this paper
we do not use any observational constraint on the degree of
triaxiality of the Milky Way DM halo.

The effects of triaxiality on the estimates for the local DM density
have been studied in Ref.~\cite{Zemp:2008gw}, using a simulation with
only DM, and in Ref.~\cite{Pato:2010yq}, with simulations of two
Milky-Way-like galaxies including baryons.  In the DM-only simulation, 
a maximum total spread of a factor of $\pm 2$ with a standard deviation
$\sigma(\bar \rho)/\bar \rho \approx 0.26$ at 8~kpc from the galactic
center was found~\cite{Zemp:2008gw}.  With the inclusion of baryons,
it was concluded that halo asphericity could lead up to $41\%$
overdensity at the local distance with respect to the spherically
averaged value~\cite{Pato:2010yq}. The impact of
halo asphericity on the so-called $J$ factors in indirect searches is
briefly discussed in Ref.~\cite{Athanassoula:2005dw}.  

In this work, we present an analysis based on a sample of $\sim 10^5$
DM-only halos from the \verb"Bolshoi" simulation~\cite{Klypin:2010qw}.
The selected halos are compatible with the mass range of the Milky
Way.  Using this sample we construct the probability distributions of
the parameters that define the density profile.  The goal of this work
is to study the impact of our lack of knowledge about the triaxial
nature of the Milky Way DM halo on both direct and indirect searches
of DM and hence, to estimate the amplitude of the systematic
uncertainties originating from the halo asphericity in the
determination of the local DM density and on the $J$ factors.

The outline of the paper is as follows.  In Sec.~\ref{sec:DDID} we
emphasize the dependence of direct and indirect DM detection searches
on the local DM density and on the $J$ factors.  In
Sec.~\ref{sec:simulations}, we discuss the data set we select from the
\verb"Bolshoi" simulation and the parametrization of the triaxial
density profile for all the halos.  In Sec.~\ref{sec:impact}, we use
three examples to illustrate the impact of halo asphericity on the
departure of the local DM density and $J$ factors from their respective
spherically averaged values.   In Sec.~\ref{sec:priors}, we describe
the different observational  constraints we consider.  In
Sec.~\ref{sec:results}, using the whole data set of halos, we present
our results for the overall probability distributions of the deviations
from the spherical averages and discuss the uncertainties owing to 
triaxiality on the local DM density and the $J$ factors. Finally,
Sec.~\ref{sec:concl} is devoted to the discussion of our results and to
the conclusion.

\section[Direct and indirect dark matter searches]{Direct and indirect
  DM searches} 
\label{sec:DDID}

Direct detection experiments are designed to detect DM particles
through their scattering with the nuclei in the detector, by
measuring low energy nuclear recoils.  The rate of such events are
proportional to the flux of DM particles streaming through the surface
of the Earth, which is in turn proportional to the local DM density,
$\rhosun$. Qualitatively, the event rate R is given
by~\cite{Bertone:2004pz}  
\begin{equation}
\label{eq:DDrate}
R \approx \frac{\rhosun \, \sigma \left< v \right>}{m_\chi \, m_A} ~,
\end{equation}
where $\sigma$ is the scattering cross section, $\left< v \right>$
is the average relative speed of DM with respect to the target,
$m_\chi$ is the DM mass and $m_A$ is the mass of the target nuclei.
Thus, when measuring recoil rates, the local DM density is fully
correlated with the scattering cross section. Hence, accurate
determination of the local DM density is crucial to 
constraint particle physics models.

In addition to direct searches, the local DM density is of key
importance in estimating the number of neutrino events from DM
annihilations accumulated in the Sun. The flux of neutrinos at the
detectors on Earth is proportional to the annihilation rate of DM
particles in the Sun, which is proportional to the Sun's capture rate 
of DM particles, which in turn is proportional to the flux of DM
particles through the Sun, and hence it is proportional to the local
DM density.  

There are two methods to estimate the value of the local DM density.
Either by constructing three-component (disc, bulge and dark
halo) models for the galaxy and confronting them against observational
data~\cite{Caldwell:1981rj}, or by calculating the DM density locally
from the stellar kinematics in our neighborhood~\cite{Bahcall:1983jb,
  Kuijken:1989a, Kuijken:1989hu, Bahcall:1991qs}.  The analyses of
the first type assume axisymmetric components and the value of $\rhosun$
obtained from them actually refers to the spherically averaged density
$\langle \rhosun \rangle$.  The value that has been usually assumed,
within an uncertainty of a factor of two, is $\langle \rhosun \rangle
\simeq 0.3$~GeV/cm$^3$.  Several recent analyses have been performed
using new data.  Some of them model the galaxy and use a large and new
set of observational constraints of the Milky Way~\cite{Catena:2009mf,
  Weber:2009pt, Strigari:2009zb, deBoer:2010eh, Iocco:2011jz,
  McMillan:2011wd, Nesti:2013uwa}, whereas in
Ref.~\cite{Salucci:2010qr} no modeling of the galaxy was needed.  

Studies of the  kinematics of stars around the Sun have also provided
recent estimates, which are in general in agreement with the previous
studies~\cite{Holmberg:1998xu, Holmberg:2004fj, Bovy:2012tw,
  Zhang:2012rsb, Bovy:2013raa}, although with larger errors.  Some of
these works have even suggested a local DM density as large as $\sim
1$~GeV/cm$^3$~\cite{Garbari:2011dh, Garbari:2012ff}.  A thorough
review of all the local DM density measurements can be found in
Ref.~\cite{Read:2014qva}, where a compilation of data from the
literature is presented and values for the spherical estimates are 
found to fall into the range $\sim 0.2$-$0.6$~GeV/cm$^3$.

As for indirect detection of DM annihilations or decays from the
galactic center, the halo shape is also important.  This can be seen as 
follows.  The differential flux of prompt gamma-rays and neutrinos
generated from DM annihilations (decays) in the smooth Milky Way DM
halo after the hadronization, fragmentation and decay of the final
states and coming from a direction within a solid angle
$\Delta\Omega$, can be written as~\cite{Bergstrom:1997fj}  
\begin{eqnarray}
\label{eq:flux}
\frac{d\Phi_{\rm ann}}{dE} (E, \Delta\Omega) & = &
\frac{\langle\sigma v\rangle}{2\,m_\chi^2} \, \sum_i {\rm BR}_i \,
\frac{dN^i_{\rm ann}}{dE}\, \bar{J}_{\rm ann}(\Omega)
\,\frac{\Delta\Omega}{4\,\pi} ~, \nonumber \\ 
\frac{d\Phi_{\rm dec}}{dE} (E, \Delta\Omega) & = &
\frac{1}{m_\chi \, \tau_\chi} \, \sum_i {\rm BR}_i \,
\frac{dN^i_{\rm dec}}{dE}\, \bar{J}_{\rm dec}(\Omega)
\,\frac{\Delta\Omega}{4\,\pi} ~,
\end{eqnarray}
where $\langle\sigma v\rangle$ is the thermal average of the total DM
annihilation cross section times the relative velocity, $\tau_\chi$ is
the DM lifetime, the discrete sum is over all DM annihilation (decay)
channels, BR$_i$ is the branching ratio of DM annihilation (decay)
into the $i$-th final state and $dN^i_{\rm ann}/dE$ ($dN^i_{\rm
  dec}/dE$) is the differential gamma-ray or neutrino spectrum for the
$i$-th channel.  The quantities $\bar{J}_{\rm ann}(\Omega)$ and
$\bar{J}_{\rm dec}(\Omega)$, which depend crucially on the DM
distribution are defined as  
\begin{eqnarray}
\label{eq:J}
\bar{J}_{\rm ann}(\Omega) & = & \frac{1}{\Delta\Omega} \,
\int_{\Delta\Omega} \dd \Omega \, \int_{\rm los}
\rho\big(r(s,\Omega)\big)^2 \, \dd s ~, \nonumber \\  
\bar{J}_{\rm dec}(\Omega) & = & \frac{1}{\Delta\Omega} \,
\int_{\Delta\Omega} \dd \Omega \, \int_{\rm los}
\rho\big(r(s,\Omega)\big) \, \dd s ~,
\end{eqnarray}
where the spatial integration of $\rho(r)^2$ and $\rho(r)$ is performed
along the line of sight within the solid angle of observation
$\Delta\Omega$.  Let us stress that for DM annihilations, the $J$
factor depends on the square of the DM density, whereas for decays it
depends linearly on the DM density.  Therefore, these quantities depend,
not only on the local DM density, but also on the shape of the halo in
the direction of observation.  Indeed, gamma-ray observations could,
in turn, be used to constrain the density profile~\cite{Weber:2007pt,
Dodelson:2007gd, Bernal:2010ip, Hooper:2010mq, Bernal:2010ti,
Bernal:2011pz, Lu:2013kda, Daylan:2014rsa}.

\section{Simulations}
\label{sec:simulations}

We use the results obtained from a large N-body cosmological
simulation dubbed \verb"Bolshoi"~\cite{Klypin:2010qw}.  The data used
in this work is publicly available through the {\it MultiDark
  Database}\footnote{\texttt{http://www.multidark.org/MultiDark/MyDB}}  
presented by Ref.~\cite{Riebe:2011gp}.  The \verb"Bolshoi" simulation
follows the non-linear evolution of DM density fluctuations in a cubic
volume of length $250$\hMpc sampled with $2048^3$ particles.  The
Adaptive Refinement Tree (ART) code was used~\cite{Kravtsov:1997vm}
and a detailed description of this simulation can be found in
Ref.~\cite{Klypin:2010qw}.
 
The cosmological parameters in this simulation are compatible with the
results from the ninth year data releases from the Wilkinson
Microwave Anisotropy Probe~\cite{Hinshaw:2012aka}, with $\Omega_m=0.27$,
$\Omega_{\Lambda}=0.73$, $n_{s}=0.95$, $h=0.70$ and $\sigma_8=0.82$
for the matter density, dark energy density, slope of the matter
fluctuations, the Hubble constant at $z=0$ in units of 100 km s$^{-1}$
Mpc$^{-1}$ and the normalization of the power spectrum, respectively.
With these parameters the mass of a simulation particle is $m_p =
1.4\times 10^{8}$\hMsun.

\subsection{Halo finding}

We use halos that were defined using the Bound Density Maxima (BDM)
algorithm~\cite{Klypin:1997sk}.  The first step in the algorithm is
finding the density at the positions of the particles in the
simulation using a top-hat filter with typically 20 particles.  After 
finding all maxima, halos are defined as the spheres of radius
$r_\Delta$ (centered around the maxima) which contain an overdensity 
mass $M_{\Delta} = \frac{4\pi}{3}\,\Delta \rho_c(z)\,r_{\Delta}^{3}$,
where $\rho_c(z)$ is the critical density of the Universe at redshift
$z$ and $\Delta$ is a given overdensity, with~\cite{Bryan:1997dn}  
\begin{eqnarray}
& & M_v = \frac{4\pi}{3} \, \Delta_v \, \rho_c(z) \, r_v^{3} ~,
  \nonumber \\ 
\Delta_v & = & 18 \pi^2 + 82\,(\Omega_m(z)-1)-39\,(\Omega_m(z)-1)^2 ~.
\end{eqnarray}

Let us note that there are a few cases where the halo is truncated to
have a radius smaller than $r_v$.  These few cases correspond to halos
that are about to merge with other massive structures, so the radius
corresponds to the distance to the surface where the density raises
again due to the proximity to the soon-to-be host halo.  The particles
in the halo are also subject to an unbinding process, whereby the
kinetic energy of each particle is compared against the gravitational
potential.  Particles that are found to be gravitational unbound are
removed from the halo.

\subsection{Shape parameter}

In order to model halos as ellipsoids, the axis ratios and orientation
is obtained by diagonalizing the shape tensor (see
Ref.~\cite{Zemp:2011ed} for a critical discussion of different
definitions), computed from all the bound particles inside the halo
radius,  
\begin{equation}
\label{eq:S}
S_{ij}  = \sum_{n}\frac{x_{i,n}x_{j,n}}{r_n^2} ~,
\end{equation}
where $x_{i,n}$ is the $i$-th Cartesian coordinate and $r_n$ is the
radial coordinate of the $n$-th particle in the halo, respectively.
The eigenvalues of this tensor determine the length of the 
axes, $a \geq b \geq c$, and its eigenvectors determine the
orientation of the halo.  The axes ratios, $b/a$ and $c/a$,
correspond to ratios of eigenvalues. 

The algorithm does not include any correction due to the fact that
$S_{ij}$ is calculated on a spherical region.  However we apply such a
correction, which depends on the halo concentration, being smaller for
more concentrated halos.  Defining $\gamma = r_{\rm rms}/r_v$, where
$r_{\rm rms}=\sum_n m_n \, r_n^2/\sum_n m_n$, the true axial ratios
are computed using the following formulae~\cite{Riebe:2011gp}: 
\begin{equation}
\label{eq:shape1}
\left(\frac{c}{a}\right)_{\rm true}  =
\left(\frac{c}{a}\right)^{\alpha},\ \alpha = 1 + 2\,{\rm max}[\gamma -
0.4,\,0]  +(5.5\,{\rm max}[\gamma-0.4,\,0])^3 ~,
\end{equation} 

\begin{equation}
\label{eq:shape2}
\left(\frac{b}{a}\right)_{\rm true}  =
\left(\frac{b}{a}\right)^{\beta},\ \beta = 1 + 2\,{\rm max}[\gamma -
0.4,\,0] +\left(5.7\,{\rm max}[\gamma-0.4,\,0]\right)^3 ~.
\end{equation} 
These relationships were calibrated against the results of a more
computationally expensive algorithm for measuring halo
shapes~\cite{Allgood:2005eu}. This method starts from the spherical
analysis provided by the BDM algorithm, then, ellipsoidal
boundaries and orientation is redefined in accordance to the
determined values of $b/a$ and $c/a$, as described above; by keeping
the longest axis, $a$, equal to the radius of the spherical
region. Next, the shape tensor is recomputed using the elliptical
norm, $r_{e, n}^2 = x_n^2 + y_n^2/(b/a)^2 + z_n^2/(c/a)^2$, instead of
the radial distance $r_n$.  The diagonalization procedure is repeated
and new values of $b/a$ and $c/a$ are obtained. This procedure is
iterated until the axes ratios converge to the desired accuracy. It
should be noted that during this process, the mass inside the
ellipsoid can indeed change, so strictly speaking, the value we use
for the halo mass is not consistent with the ellipsoidal shape.
However, the differences between these two quantities are at most
$20\%$ (see, e.g., Ref.~\cite{MunozCuartas:2010ig}), which is much
less than the variation across the halo population. Therefore, we do
not implement any conversion between the halo mass inferred from the
spherical BDM algorithm and the expected ellipsoidal counterpart.

\subsection{Concentration parameter}

The concentration values reported in the {\it Multidark Database} are
computed for a spherical Navarro-Frenk-White (NFW)
profile~\cite{Navarro:1995iw, Navarro:1996gj},
\begin{equation}
\label{eq:nfw1}
\rho(r)=\frac{4\,\rho_s}{r/r_s\left(1+r/r_s\right)^2} ~,
\end{equation}
where $r$ is the distance to the center, $\rho_s$ is the density at
$r_s$, the scale radius, which corresponds to the distance at which
the logarithmic slope of the profile is $\frac{\dd \log\rho}{\dd
  \log\,r}(r_s)=-2$.

Following the algorithm of Ref.~\cite{Prada:2011jf}, the halo
concentration is defined from the ratio of the the maximum circular
velocity, $V_{\rm max}$, to the circular velocity at the halo radius,
$V_v$.  For each halo the circular velocity $V_{\rm
  circ}(r)=\sqrt{GM(r)/r}$ is computed using the radial mass
distribution $M(r)$ for all bound particles.  The separation of the
radial bins is $\Delta \log r/R_v=0.01$.   

For the case of a spherical NFW density profile, the concentration is
found by numerically solving the equation
\begin{equation}
\label{eq:ratio}
\left(\frac{V_{\rm max}}{V_v}\right)^2 = \frac{0.2162\,c_v}{f(c_v)} ~,
\end{equation}
where $f(c_v)$ is
\begin{equation}
\label{eq:c_function}
f(c_v) = \ln(1+c_v) - \frac{c_v}{1+c_v} ~.
\end{equation}
The results of this method are computationally robust, in contrast to
more uncertain radial fitting methods that strongly dependent on the
range used for the fit~\cite{Klypin:2010qw, Meneghetti:2013nma}.  The
comparison of these two methods yield an offset $<15\%$, where the
concentrations derived with the velocities are higher.  For halos with 
$c_v<5$ the offset is smaller than the intrinsic scatter at fixed halo
mass~\cite{Prada:2011jf}.

On the other hand, in order to describe a triaxial halo we use a halo
profile of the NFW form
\begin{equation}
\label{eq:nfwnsph}
\rho(r_e) = \frac{4 \, \rho_e}{c_e \frac{r_e}{r_v}\left(1 + c_e
  \frac{r_e}{r_v}\right)^2} ~,  
\end{equation}
where $c_e$ is the concentration parameter in the triaxial case and
the radial coordinate $r$ is replaced by its ellipsoidal counterpart 
\begin{equation}
\label{eq:relip}
r_e^2 = x^2 + \left(\frac{y}{b/a}\right)^2 +
\left(\frac{z}{c/a}\right)^2 ~,
\end{equation} 
where $(x,y,z)$ are Cartesian coordinates. In order
to be consistent with the halo mass inferred from the spherical BDM, 
the density at $r_e = r_v/c_e$ ($\rho_e$), is set by imposing the
following condition on the halo mass:  
\begin{equation}
\label{eq:intmv}
M_v = \displaystyle\int \dd V \rho(r_e) ~,
\end{equation}
where $\dd V$ is the spherical volume within a radius $r_v$.    

The {\it Multidark Database} provides us with the halo mass $M_v$, the
ratios of the axes, $b/a$ and $c/a$, and concentration parameter
$c_v$, obtained from Eq.~(\ref{eq:ratio}) assuming a spherical NFW
halo.  However, in order to consistently define a triaxial halo, we
need to determine the corresponding concentration parameter $c_e$.  By
numerically computing the ratio of $V_{\rm max}$ and $V_v$ for each
halo with profile $\rho(r_e)$ of Eq.~(\ref{eq:nfwnsph}) and equating
it to the right-hand side of Eq.~(\ref{eq:ratio}), we obtain $c_e$.

We have found numerically an approximate relation between $c_e$ and
$c_v$, which reads
\begin{equation}
\label{eq:celcv}
c_e \simeq \left(\frac{b}{a} \, \frac{c}{a}\right)^{1/3} \, c_v ~.
\end{equation}
The expression is accurate up to a precision of $3\%$ - $15\%$ in our
sample.  Heuristically, one can understand this relation by noting
that the radius of a sphere with the same volume as the ellipsoid is
$r_{e,v} = [(b/a) \, (c/a)]^{1/3} \, r_v$, so for the same scale
radius, $r_v/c_v = r_{e,v}/c_e$, which is exactly Eq.~(\ref{eq:celcv}).

\subsection{Data set}

\begin{figure}[t]
\centering
\begin{center}
\begin{tabular}{ll}
\includegraphics[width=0.4\textwidth,angle=0]{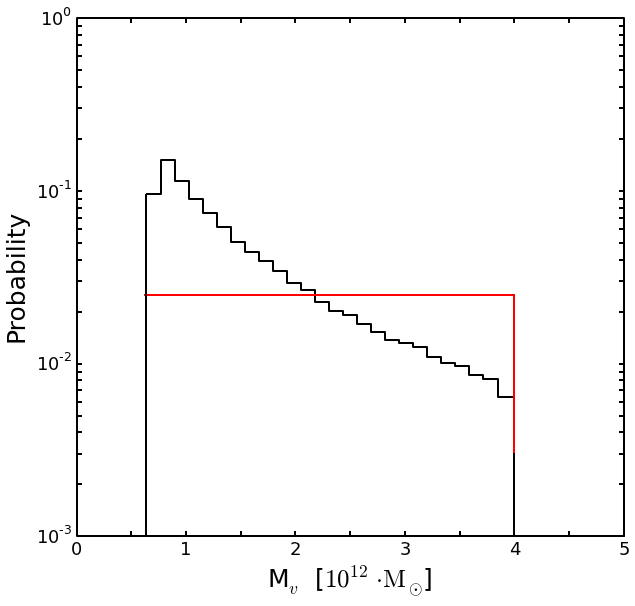} & 
\includegraphics[width=0.4\textwidth,angle=0]{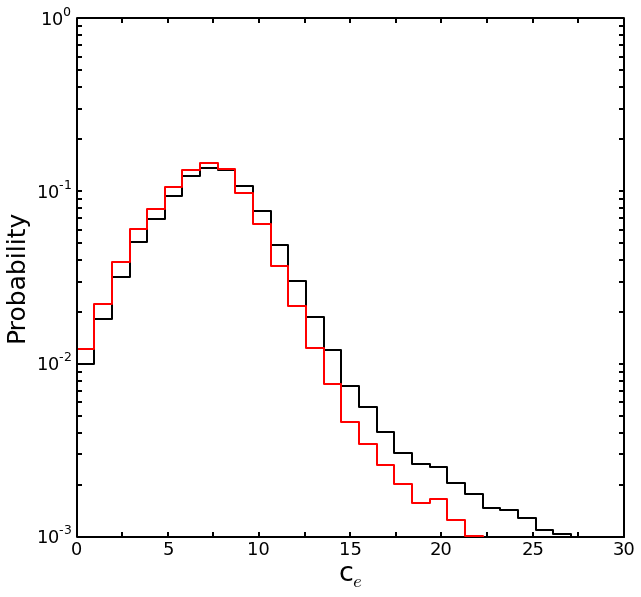} \\
\includegraphics[width=0.4\textwidth,angle=0]{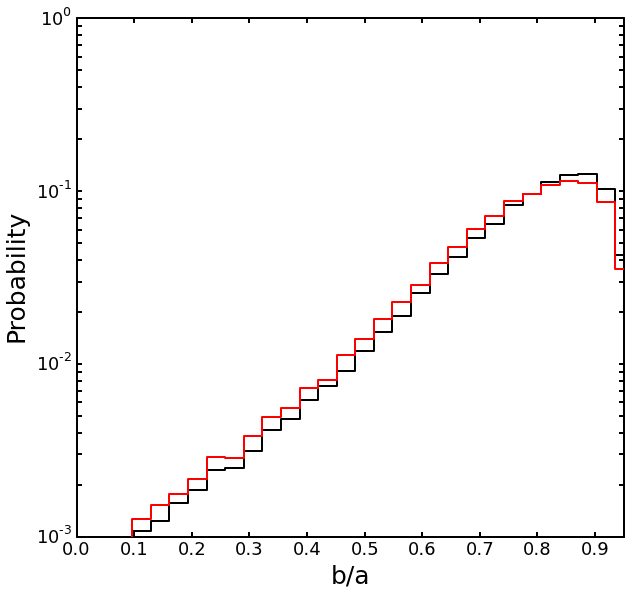}&
\includegraphics[width=0.4\textwidth,angle=0]{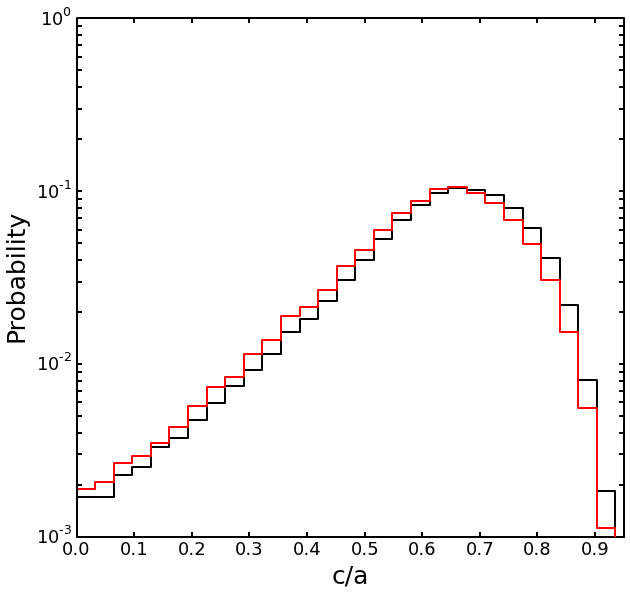} \\ 
\end{tabular}
\end{center}
\caption{\textbf{\textit{Probability distributions of the halo
      parameters}}: black lines for our final sample and red
  lines when a flat prior on $M_v$ is imposed.  The upper two
  panels depict the probability distributions of the halo mass $M_v$
  (left panel) and concentration parameter $c_e$ (right panel).  The
  lower two panels depict probability distributions of the axes ratios
  $b/a$ (left panel) and $c/a$ (right panel).}
\label{fig:pardist}
\end{figure}

The original data set contains $\sim 10^5$ halos which are selected to
have masses in the Milky Way range, i.e., $M_v = [0.7, \, 4.0] \times
10^{12} \, \Msun$.  However, the sample of halos has some spurious
elements which cannot represent the Milky Way.  For instance, as
mentioned above, there are halos that are close to other massive
structures and are truncated by the BDM algorithm.  Therefore we 
discard halos with a radius $r_v<220$~kpc, which reduces the size of
the original sample by $\sim 5\%$, thereby resulting in 87132 halos.
The resulting distributions for the parameters of triaxial NFW profile
are shown with the black lines in Fig.~\ref{fig:pardist}.  The average
values of the parameters in this sample are: $\langle M_v \rangle =
1.55\times 10^{12} \, \Msun$, $\langle c_e \rangle = 8.9$, $\langle
b/a \rangle = 0.81$ and $\langle c/a \rangle = 0.66$.

\begin{figure}[t]
\centering
    \includegraphics[width=0.5\textwidth]{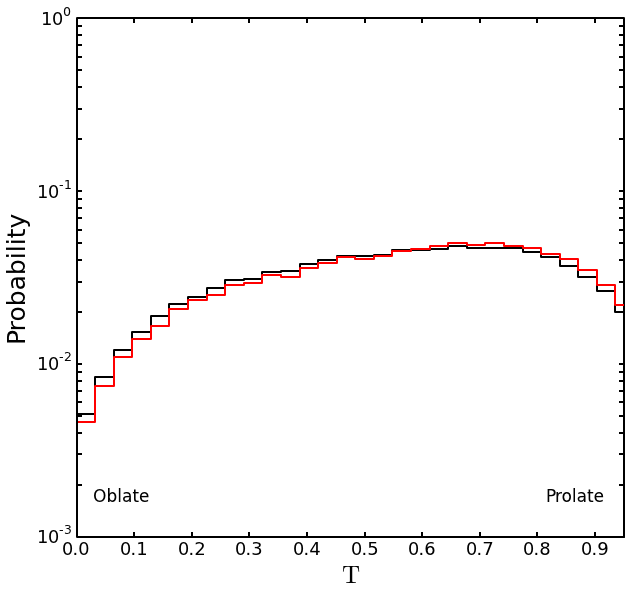}  
\caption{\textbf{\textit{Probability distribution of the triaxiality
      parameter (T)}} in the data set.  The black line
  represents our final sample and the red line represents
  the re-weighted sample with a flat prior on $M_v$.}
\label{fig:triax}
\end{figure}

A useful way to characterize the shape of an ellipsoidal halo is by
evaluating the so-called triaxiality parameter~\cite{Franx:1991}
\begin{equation}
\label{eq:T}
T=\frac{1-(b/a)^2}{1-(c/a)^2} ~.
\end{equation}
An ellipsoid is considered prolate (sausage shaped) if $a \gg b
\approx c$ ($1>T>2/3$), triaxial if $a > b > c$ ($2/3>T>1/3$) and
oblate (pancake shaped) if $a \approx b \gg c$ ($1/3>T>0$).  It is
well known that DM halos in N-body simulations which neglect possible
baryonic effects, are in general triaxial: close to prolate in the
central part and becoming rounder in the outer
parts~\cite{Frenk:1988zz, Dubinski:1991bm, Katz:1991, Warren:1992tr,
  Dubinski:1993df, Jing:2002np, Kasun:2004zb, Bailin:2004wu,
  Hopkins:2004np, Allgood:2005eu, Maccio':2006nu, Bett:2006zy,
  Hayashi:2006es, Kuhlen:2007ku, Stadel:2008pn, MunozCuartas:2010ig,
  VeraCiro:2011nb, Schneider:2011ed, Vera-Ciro:2014ita}.  Numerical
simulations with baryons produce shapes closer to
spherical~\cite{Kazantzidis:2004vu, Bailin:2005xq, Gustafsson:2006gr,
  Kazantzidis:2010jp, Bryan:2012mw}.  We show in Fig.~\ref{fig:triax}
the distribution of the triaxial parameter $T$ in our final sample,
which has an average value of $\langle T \rangle = 0.58$.

Hierarchical structure formation predicts that massive halos are
formed by mergers of smaller halos.  This feature is nicely
illustrated in the probability distribution of $M_v$ in
Fig.~\ref{fig:pardist}.  A larger number of halos exist for low
masses.  This dependence of the halo abundance as a function of halo 
mass is well understood and is parametrized through the halo mass
function.  However, as for the Milky Way, its mass is quite uncertain
(within an order of magnitude).  In order to avoid an unfair weight to  
the low mass range due to cosmological effects in the simulation, we
also consider a flat prior on $M_v$ by weighting all bins such that all
values of $M_v$, in the range $M_v = [0.7, \, 4.0] \times 10^{12} \,
\Msun$, are equiprobable.
 With such an exercise we can study
systematic effects independently of the cosmological bias. The
probability distributions of the parameters for this re-weighted
sample, with a flat prior on the halo mass, are depicted in
Figs.~\ref{fig:pardist} and~\ref{fig:triax} with the red 
curves.

\section{Impact of halo asphericity}
\label{sec:impact}

In this section we describe the possible impact of having a
non-spherical halo on quantities such as the local DM density
($\rhosun$) and $J$ factors.  Consider a set of observers at a given
distance from the halo center who are able to locally measure
properties of the halo and are able to compute $\rhosun$.  In a
spherically symmetric halo all observers would obtain the same value. 
However, in a triaxial halo these measurements would lead to a
significant variance with respect to the spherically averaged
value~\cite{Knebe:2006hh}.  We show that there exists similar
deviations from the spherically averaged value for $J$ factors as
well.

In order to illustrate this, we consider three example halos: an
almost spherical halo, a prolate halo and an oblate halo; which are
described by the parameters in Table~\ref{t:impact-asph}.  A typical
ellipsoid is defined by three axes of symmetry, the major axis ($a$),
intermediate axis ($b$) and the minor axis ($c$).  Correspondingly
there exists three orthogonal planes denoted by $a-b$, $a-c$ and
$b-c$.  We assume the galactic disc to coincide with one of the
symmetry planes of the dark halo, although this is not
guaranteed~\cite{Debattista:2013rm, Deg:2014yla}.  However, due to the
lack of a definite solution to the problem of the shape and
orientation of the halo, we assume the alignment of the stellar disc
with one of the three symmetry planes of the halo in order to bracket
the uncertainties.  Consequently, our ignorance of where the solar
system might reside in a triaxial halo motivates us to evaluate the
aforementioned quantities individually for each plane of symmetry.

\begin{figure}[t]
\centering
\begin{center}
\begin{tabular}{ccc}
\qquad {\bf Approx. Spherical} & \qquad {\bf Prolate} & \qquad {\bf
  Oblate}\\ 
\includegraphics[width=0.3\textwidth,angle=0]{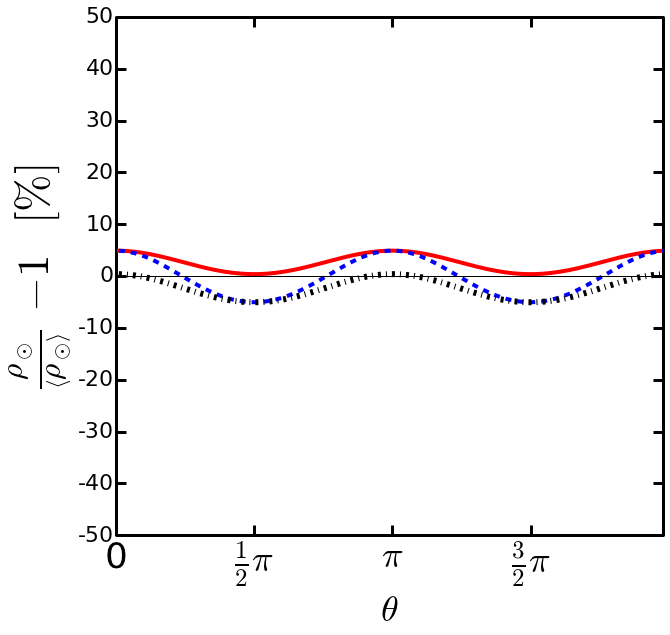} &
\includegraphics[width=0.3\textwidth,angle=0]{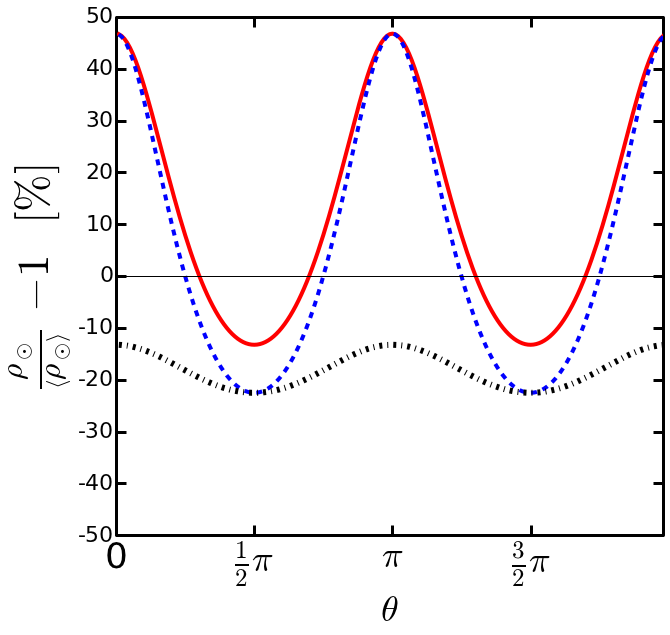}&
\includegraphics[width=0.3\textwidth,angle=0]{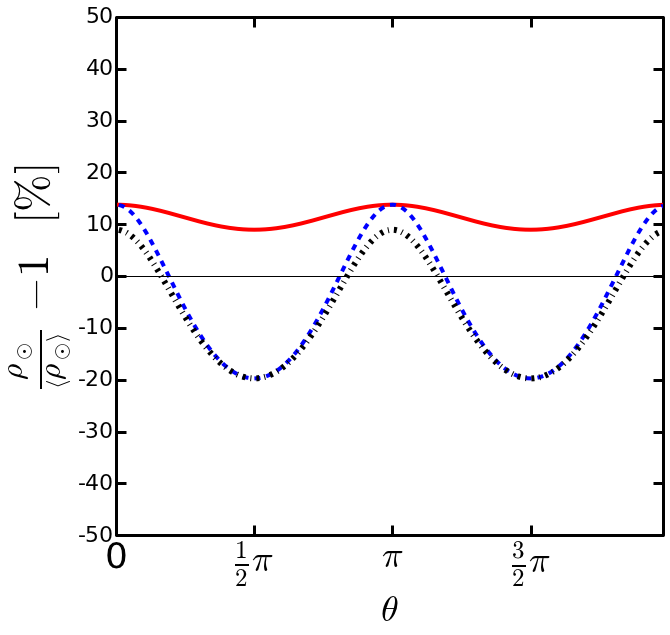}
\\ 
\includegraphics[width=0.3\textwidth,angle=0]{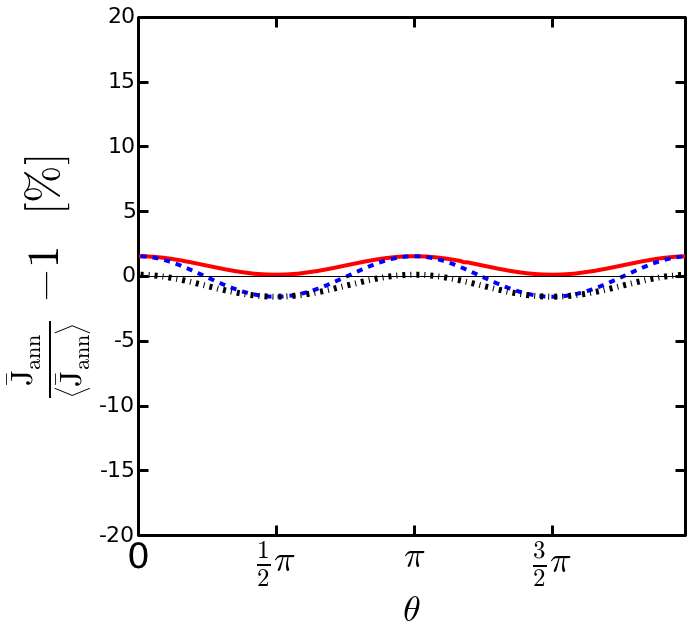}&
\includegraphics[width=0.3\textwidth,angle=0]{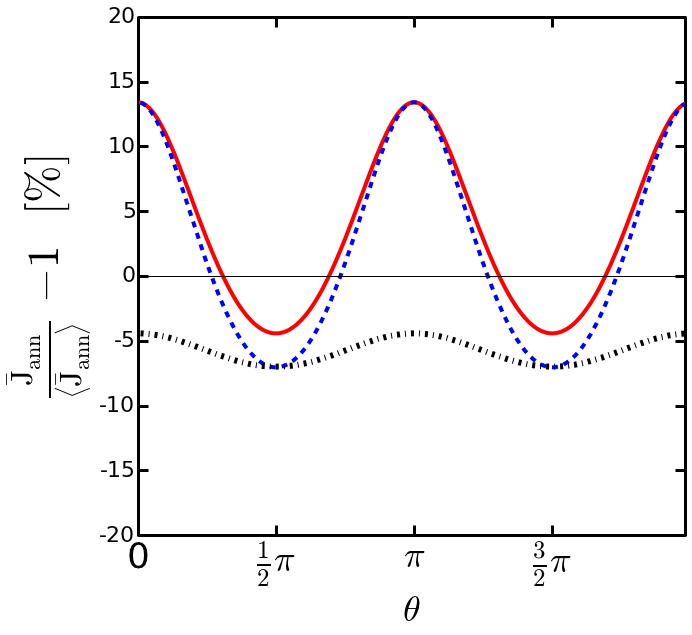}& 
\includegraphics[width=0.3\textwidth,angle=0]{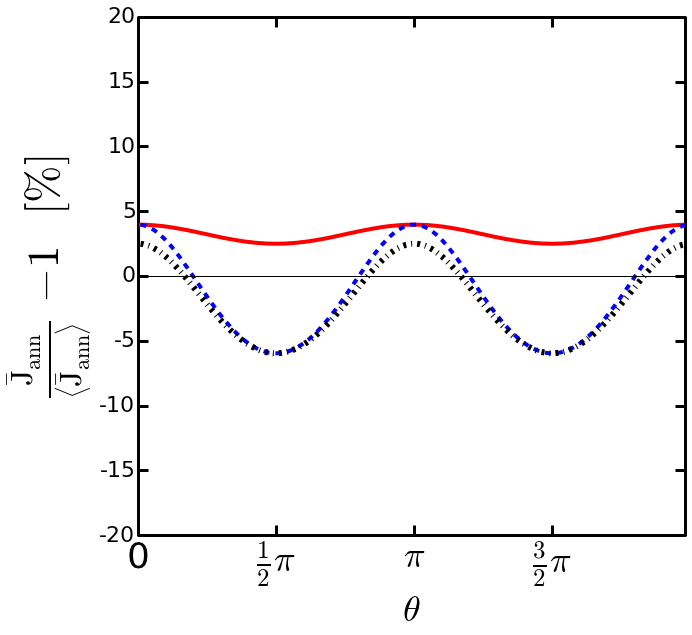}\\ 
\includegraphics[width=0.3\textwidth,angle=0]{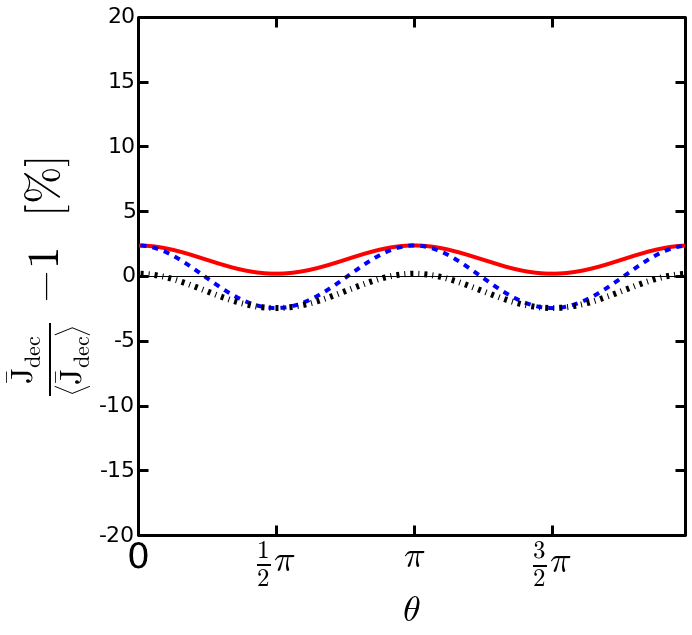}&
\includegraphics[width=0.3\textwidth,angle=0]{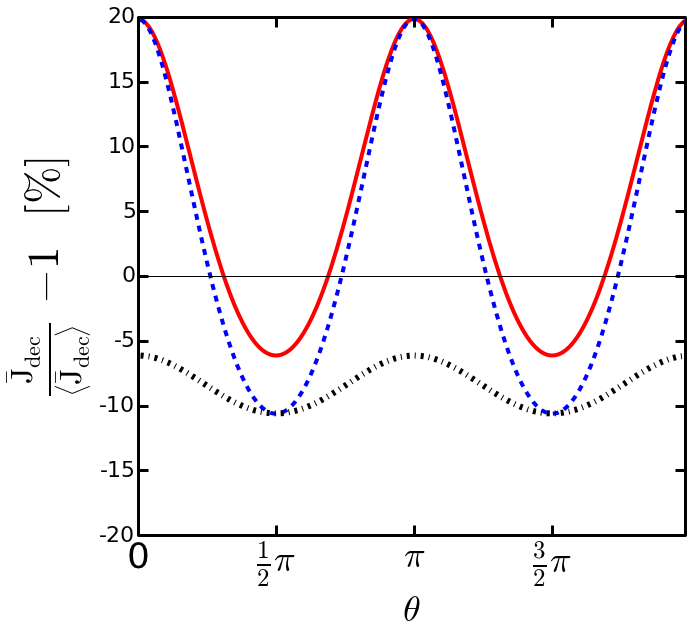}& 
\includegraphics[width=0.3\textwidth,angle=0]{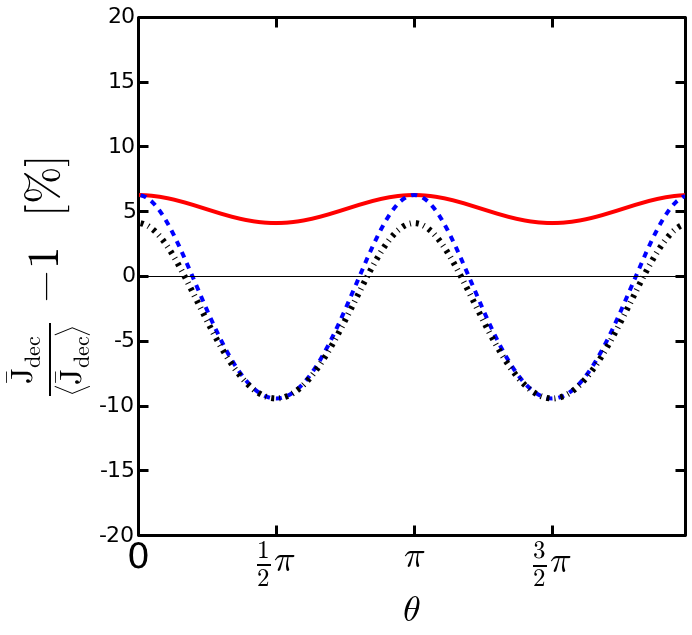}
\end{tabular}
\end{center}
\caption{\textbf{\textit{Deviation from the average in a spherical
      shell of radius \boldmath{$\mathit{R}_{\odot}$} for the three
      symmetry axes.}} We show the results for three quantities:
  $\rhosun$ (top panels), $\bar{J}_{\rm ann}$ (middle panels) and
  $\bar{J}_{\rm dec}$ (bottom panels) for a square ROI of
  $3^{\circ}\times 3^{\circ}$ around the galactic center.  Each column
  refers to a given halo: approximately spherical halo (left column),
  prolate halo (middle column) and oblate halo (right column).  In all
  panels, the $a-b$ plane is represented by the solid red curve, the
  $a-c$ plane by the dashed blue curve and the $b-c$ plane by the
  dotted black curve.  The angle $\theta$ represents the angular
  position along a circle of radius $\Rsun$ for each plane of symmetry.}
\label{fig:haloimpact}
\end{figure}

With the halo profile exactly defined by $M_v$, $c_e$, $b/a$ and
$c/a$, we proceed to compute the local DM density, $\rhosun$, and the
$J$ factors for a region of interest (ROI) of $3^{\circ}\times
3^{\circ}$ (a square of $3^{\circ}$ side) around the galactic center. 
We do so for different points along a circle of radius  $\Rsun$ (= 8.3
kpc) for the three planes of symmetry.  Then, we compute the average
quantities in a spherical shell of the same radius\footnote{The
  average density $\langle \rhosun \rangle$ is the quantity inferred
  from dynamical measurements in the galaxy.}, $\langle \rhosun
\rangle$, $\langle \bar{J}_{\rm ann} \rangle$ and $\langle
\bar{J}_{\rm dec} \rangle$.  The results corresponding to the three
halos (one approximate spherical, one prolate and one oblate) defined in
Table~\ref{t:impact-asph} are depicted in Fig.~\ref{fig:haloimpact},
where we show the deviation of each quantity with respect to its 
spherical average as a function of the angular position $\theta$ along
a circle of radius $\Rsun$.  We choose $\theta=0$ as a reference
point, which corresponds to the occurrence of the maximum value of 
each quantity.  In all panels, the solid red, dashed blue and dotted
black curves indicate the variation along the $a-b$ plane, the $a-c$
plane and the $b-c$ plane, respectively.

\begin{table}[t]
\begin{center}
\renewcommand\arraystretch{2}
\begin{tabular}{|c||c|c|c|c|c|}
\hline
Halo Type &  $M_v$ [$ 10^{12} \, \Msun$] & $R_v$ [kpc]&$c_e$ & $b/a$ &
$c/a$ \\ 
\hline\hline
Approx. Spherical & 3.8 & 242 & 9.73 & 0.97 & 0.91\\
\hline 
Prolate           & 3.6 & 404 & 5.33 & 0.58 & 0.48\\
\hline 
Oblate            & 2.0 & 419 & 9.79 & 0.97 & 0.77\\
\hline 
\end{tabular}
\caption{Parameters of the example halos used to illustrate the impact
  of asphericity in Fig.~\ref{fig:haloimpact}.} 
\label{t:impact-asph}
\end{center}
\end{table}

As expected, the deviations from the spherical average are the
smallest for the approximately spherical halo (left panels in 
Fig.~\ref{fig:haloimpact}).  A maximum variation of $\sim 5\%$ for the
local DM density is found and only $\sim 1.6\%$ and $\sim 2.5\%$ for
$\bar{J}_{\rm ann}$ and $\bar{J}_{\rm dec}$, respectively.  On the
other hand, in the case of the prolate halo we consider here (middle
panels in Fig.~\ref{fig:haloimpact}), deviations of up to $\sim 46\%$,
$\sim 14\%$ and $\sim 20\%$ are possible for $\rhosun$, $\bar{J}_{\rm
  ann}$ and $\bar{J}_{\rm dec}$, respectively.  The oblate halo (right
panels in Fig.~\ref{fig:haloimpact}), being closer to the spherical
case, presents deviations of up to $\sim 20\%$, $\sim 6\%$ and $\sim
9\%$ for $\rhosun$, $\bar{J}_{\rm ann}$ and $\bar{J}_{\rm dec}$,
respectively. 
 
A common feature among the different quantities shown in
Fig.~\ref{fig:haloimpact} is their angular dependence, i.e., the
peaks and troughs occur at the same angular position.  This is related
to the exact position of observation on a given plane.  As we move
around a circle of radius $\Rsun$ from the galactic center, the
angular dependence is periodic.  The amplitude is proportional to the
overall normalization, which in turn depends on the shape of the halo.
For example, oblate halos have larger deviations from the spherical
average on the $b-c$ plane than on the $a-b$ plane.  This trend is
inverted for a prolate halo.  Let us also note that the deviations for
the $J$ factors are larger for DM decay than for DM annihilation.
Heuristically, this can be understood by considering the contributions
to the $J$ factors away from the center of the halo.  As the flux from
DM annihilations depends on the square of the DM density distribution,
in contrast to the linear dependence for DM decays. For DM
annihilations, the relative contribution to the $J$ factors from the
outer regions (i.e., regions which are closer to the boundary of the
ROI) with respect to the contribution from the center is expected to
be smaller than for DM decays.  Hence, the deviations from the
spherical average are smaller for DM annihilations.  This is seen in
Fig.~\ref{fig:haloimpact} and below when discussing our results. 

The point emphasized by this exercise is that halo asphericity could
give rise to significant deviations from the spherically averaged
values of relevant quantities for DM searches.  Indeed, these
deviations could be quite large depending on the shape of the halo and
could have a substantial impact on direct and indirect DM detection.
In the following, we quantify these uncertainties statistically by
using the whole halo data set.

\section{Observational priors}
\label{sec:priors}

In addition to showing results for the original sample from the {\it
  MultiDark Database}, which represents a selection of halos in a mass
range compatible with the mass of the Milky Way, we also consider
several observational constraints and show results after applying the 
corresponding priors.  In addition to a prior on the virial mass, we
also include priors on the enclosed mass at 60~kpc, the local DM
surface density and the Sun's galactocentric distance.  We add flat 
priors for the virial mass (to compensate for the cosmological bias)
and the Sun's distance to the galactic center. Gaussian priors are used 
for the other two observables.  Hence, for each plane of symmetry, the
probability distribution function of the original data sample is
modified as:
\begin{eqnarray}
\label{eq:priors}
{\rm PDF}_{\rm prior}^p (\vec \omega) & = & C \, \frac{{\rm PDF} (\vec
  \omega)}{{\rm PDF} (M_v)} \times
\theta (M_v - M_v^{\rm min}) \, \theta (M_v^{\rm max} - M_v)
\nonumber \\ 
& & \times \int_{\Rsun^{\rm min}}^{\Rsun^{\rm max}} \dd \Rsun
\exp\left[-\frac{(M_{60}^{\rm DM}-M_{60}
    )^2}{2\,\sigma_{60}^2}\right] \nonumber \\
& & \times \int_0^{2 \pi} \dd \psi \, 
\exp\left[-\frac{(\Sigma_{1.1}^{\rm DM} - \Sigma_{1.1}^p
    (\Rsun, \psi))^2}{2\,\sigma_\Sigma^2}\right] ~,
\end{eqnarray}
where $C$ is a normalizing constant, $\vec \omega = (M_v, c_e, b/a,
c/a)$, ${\rm PDF} (\vec \omega)$ is the original probability
distribution function and ${\rm PDF} (M_v)$ is the probability
distribution function after marginalizing over $(c_e, b/a, c/a)$.
${\rm PDF}_{\rm prior}^p (\vec \omega)$ is computed for the three
symmetry planes, $p = a-b$, $a-c$ and $b-c$, where $\psi$ is the
azimuthal angle at the solar circle.  The limits, central values and
errors, discussed below, are indicated in Tab.~\ref{tab:priors}.  Note
that the prior on $M_{60}$ is a global prior for a given halo, 
whereas the prior on the DM surface density is a local
constraint which depends on the exact position of the observer in the
halo and thus, on the plane of symmetry under consideration.  Given a
triaxial halo, the value of the surface density at various angular
points can be significantly different.

\begin{table}[t]
\begin{center}
\renewcommand\arraystretch{2}
\begin{tabular}{|c||c|c|c|c|}
\hline
 & \multicolumn{2}{ c| }{Gaussian priors} &
\multicolumn{2}{ |c| }{Flat priors} \cr
\cline{2-5}
 & Central value & $1\sigma$ error & Lower cut & Upper cut \cr 
\hline \hline 
\small{Virial mass $[10^{12} \, \Msun]$} & -- & -- &
$M_v^{\rm min} = 0.7$ & $M_v^{\rm max} = 4.0$ \cr \hline
\small{DM mass within 60~kpc $[10^{11} \, \Msun]$} & $M_{60}^{\rm DM}
= 4.0$ & 
$\sigma_{60} = 0.7$ & -- & -- \cr \hline
\small{Local DM surface density $[\Msun \, {\rm pc}^{-2}]$} &
$\Sigma_{1.1}^{\rm DM} = 17$ & $\sigma_\Sigma = 6$ & -- & -- \cr \hline
\small{Sun's galactocentric distance [kpc]} & -- & -- & $\Rsun^{\rm
  min} = 7.5$ & $\Rsun^{\rm max} = 9$ \cr \hline 
\end{tabular}
\caption{Limits for the halo virial mass ($M_v$) and the Sun's
  galactocentric distance ($\Rsun$) and central values and $1\sigma$
  errors for the DM halo mass at 60~kpc ($M_{60}^{\rm DM}$) and the local
  DM surface density ($\Sigma_{1.1}^{\rm DM}$), which are used for the
  priors discussed in the text.}
\label{tab:priors}
\end{center}
\end{table}

\subsection{The virial mass of the Milky Way}
\label{subsec:Mvir}

We have selected our data set by the criterion of halo mass (see
Sec.~\ref{sec:simulations}).  Observationally, different methods have 
been used to determine the mass of the Milky Way, such as
gravitational lensing, gas rotation curves, escape velocity arguments 
or Jeans modeling of the radial density and velocity dispersion
profiles of kinematic tracers (blue horizontal branch stars, carbon
stars, asymptotic giant branch stars, globular clusters, satellite
galaxies), or timing arguments.  In general, estimates based on stellar
kinematics tend to yield a low mass, $\lesssim 10^{12} \,
\Msun$~\cite{Smith:2006ym, Xue:2008se, Deason:2012ky, Bovy:2012ba,
  Kafle:2012az}, but usually infer the total mass from an
extrapolation from the inner halo to the virial radius using models 
for the different components of the Milky Way. However, 
estimates based on distant tracers, such as globular clusters or
satellite galaxies, and on statistics of cosmological DM simulations
tend to imply larger masses, $\gtrsim 10^{12} \,
\Msun$~\cite{Kochanek:1995xv, Wilkinson:1999hf, Li:2007eg,
  Watkins:2010fe, BoylanKolchin:2010ck, Busha:2010sg, Sohn:2012xt,  
  BoylanKolchin:2012xy}, but there are few distant tracers beyond
$\sim 80$~kpc, where the DM dominates.  There are, though, exceptions
in both cases~\cite{Kallivayalil:2009wc, Gnedin:2010fv,
  Przybilla:2010gd, 2011A&A...531A..82S, VeraCiro:2012na,
  Piffl:2013mla} and combinations of these data sets to obtain 
rotation curves up to large distances allow for a wide range of
values based on modeling~\cite{Sakamoto:2002zr, Battaglia:2005rj,
  Dehnen:2006cm, Sofue:2008wt, Sofue:2008wu, Sofue:2011kw,
  Bhattacharjee:2013exa}.  Fitting of dynamical models with kinematic
and photometric data tends to provide best fit values above $10^{12} \,
\Msun$~\cite{Klypin:2001xu, Catena:2009mf, Weber:2009pt,
  McMillan:2011wd, Nesti:2013uwa}. On the other hand, more indirect  
determinations can be obtained from combinations of galaxy-galaxy
lensing and Tully-Fisher data~\cite{Dutton:2010qa, Reyes:2011vc} or
from the relation of the halo mass and the stellar
mass~\cite{Guo:2009fn, Reddick:2012qy, Moster:2012fv}, both predicting
a mass of the Milky Way above $10^{12} \, \Msun$.  From all these
results, the virial mass of the Milky Way is expected to lie within a
large range of values, $8 \times 10^{11} < M_v/\Msun < 3 \times
10^{12}$.  In this work we have selected a data set from the {\it
  MultiDark Database}, which covers the mass interval $M_v = [0.7, 
  \, 4.0] \times 10^{12} \, \Msun$ and, as discussed in 
Sec.~\ref{sec:simulations}, we use a flat prior for this interval in 
order to compensate the cosmological bias which favors low mass halos.

\subsection{The mass of the Milky Way within 60~kpc}
\label{subsec:M60}

Within the innermost $\sim$80~kpc of the Milky Way there are
numerous kinematic tracers, beyond this distance there are only a few
known globular clusters and satellite galaxies (see
Ref.~\cite{Deason:2012ky}, for instance).  This introduces significant
errors when estimating the total mass of the Milky Way.  Therefore, we
expect the estimate of the mass in the inner $\sim 50-80$~kpc to
suffer from fewer uncertainties.  This mass has been determined using 
kinematic data of halo stars~\cite{Kochanek:1995xv, Wilkinson:1999hf,
  Xue:2008se, Gnedin:2010fv, 2011A&A...531A..82S, Deason:2012ky,
  Kafle:2012az}.  In this work we consider the result based on a
set of 2401 blue horizontal-branch halo stars from the Sloan Digital
Sky Survey with distances from the galactic center up to
$\sim$60~kpc~\cite{Xue:2008se}, $M_{60} \equiv M (<60~{\rm kpc}) =
(4.0 \pm 0.7) \times 10^{11} \, \Msun$, and implement Gaussian priors. 
Let us note that $M_{60}$ is the total mass within 60~kpc and that the
galactic disc and bulge (visible matter) are estimated to contribute
with a total mass of about an order of magnitude lower, $\sim~(5-7)
\times 10^{10} \, \Msun$~\cite{Gerhard:2002ud, Flynn:2006tm,
  McMillan:2011wd, Bovy:2013raa}, which represents approximately the
$1\sigma$ error on $M_{60}$.  We do not correct for this difference,
i.e., we take $M_{60}^{\rm DM} = M_{60}$.

\subsection[Local dark matter surface density]{Local DM surface density}
\label{subsec:SD}

The mass distribution of the Milky Way in the local neighborhood can
be characterized by the local surface density to some distance
$|z|=z_0$, 
\begin{equation}
\label{eq:Sigma}
\Sigma_{z_0} (\Rsun) \equiv \Sigma(\Rsun, |z|<z_0) =
\int_{-z_0}^{+z_0} \rho\left(\Rsun, z\right) \, \dd z ~,  
\end{equation}
where the integration limit is conventionally taken to be
$z_0=1.1$~kpc.

From the determination of the vertical gravitational potential using
stellar dynamics, the total local surface density to $|z| = 1.1$~kpc
has been inferred to lie within $\Sigma_{1.1} (\Rsun) \simeq 60-80 \,
\Msun\,{\rm pc}^{-2}$~\cite{Kuijken:1990cb, Siebert:2002nx,
  Korchagin:2003yk, Holmberg:2004fj, Bienayme:2005py, Zhang:2012rsb,
  Bovy:2013raa}.  However, establishing which fraction
belongs to the dark halo or to the baryonic disc (stars and gas)
requires further modeling.  The local disc surface density is found
to be $\Sigma_{1.1}^d (\Rsun) \simeq 45-55 \, \Msun\,{\rm
  pc}^{-2}$~\cite{Kuijken:1989a, Kuijken:1989hu, Kuijken:1989c,
  Kuijken:1990cb, Flynn:1994, Holmberg:2004fj, Garbari:2011dh,
  Garbari:2012ff, Zhang:2012rsb, Bovy:2013raa}.  Unlike dynamical
determinations of the total disc surface density, direct observations
of local stars are expected to induce smaller uncertainties on the
contribution from the visible stellar matter, $\Sigma_{1.1}^{\rm
visible} \sim 25-30 \, \Msun\,{\rm pc}^{-2}$~\cite{Gould:1995tf,
  Holmberg:1998xu, Zheng:2001wc, Binney:2001wu, Flynn:2006tm,
  Bovy:2011zx}.  An additional contribution of $\sim 5-7 \,
\Msun\,{\rm pc}^{-2}$ is estimated to come from stellar remnants and
brown dwarfs~\cite{Holmberg:2004fj, Flynn:2006tm}, and $\sim 13-15 \,
\Msun \,{\rm pc}^{-2}$~\cite{Holmberg:1998xu, Olling:2001yt} from the
interstellar gas.

We use the recent results of Ref.~\cite{Bovy:2013raa}, obtained after
dynamical modeling and based on a large data set of phase-space of
individual stars.  The total local surface density is determined to
be $\Sigma_{1.1}(\Rsun) = (68 \pm 4) \, \Msun \, {\rm pc}^{-2}$, of
which $(38 \pm 4) \, \Msun \, {\rm pc}^{-2}$ is contributed by stars
and stellar remnants.  A contribution of $13 \, \Msun \, {\rm pc}^{-2}$
from the thin gas layer is assumed. Thus implying a DM halo
contribution $\Sigma_{1.1}^{\rm DM} (\Rsun) = (17 \pm 6) \, \Msun \,
\text{pc}^{-2}$.

\subsection{Sun's galactocentric distance}
\label{subsec:Rsun}

Based on an old recommendation of the International Astronomical
Union, the Sun's galactocentric distance is usually assumed to be
$\Rsun = 8.5$~kpc~\cite{Kerr:1986hz}.  This number was obtained as a
result of an average of different estimates after 1974.  However, a
more careful analysis of estimates between 1974 and 1993 rendered a
lower value, $\Rsun = 8.0 \pm 0.5$~kpc~\cite{Reid:1993fx} and with new
measurements, about a decade  ago, the average value was $\Rsun = 7.9
\pm 0.2$~kpc~\cite{Nikiforov:2004} (see also
Ref.~\cite{Avedisova:2005}). 

In recent years, many new measurements have been published.  Different
methods exist to determine $\Rsun$ (see Ref.~\cite{Francis:2013zna}
for a compilation of $\Rsun$ measurements since 1918).  One of them is
the halo centroid method, based on the mean distance of globular
clusters~\cite{Bica:2005en, Francis:2013zna}, which tends to give low
values, $\Rsun \sim 7.2-7.4$~kpc.  Another method uses the luminosity
distance of bulge stars, such as RR Lyrae, Cepheids, Delta Scuti, the
red clump, Mira stars or planetary nebula.  The measured value varies
over a relatively large range, $\Rsun \sim
7.4-8.8$~kpc~\cite{Collinge:2006km, Nishiyama:2006zr,
  Groenewegen:2008hz, Feast:2008dw, Kunder:2008ig,
  Vanhollebeke:2009ka, Majaess:2009xc, Dambis:2009, Matsunaga:2009xq,
  Majaess:2010zu, Pietrukowicz:2011er, Nataf:2012pk, Matsunaga:2012qj, 
Francis:2013zna}, probably caused by systematics associated to the
calibration of magnitudes~\cite{Francis:2013zna}.  The solar-circle
(or its modification, the near-solar circle) method is a geometrical
method which requires no other assumption but circular
motion~\cite{Ando:2011, Sofue:2011hz, Bobylev:2012bi} and obtains a
Sun's galactocentric distance in the range $\Rsun \sim 7.3-7.8$~kpc.
Other pure geometrical approaches obtain $\Rsun \sim
8.3$~kpc~\cite{Sato:2010dr, Schonrich:2012qz}.  In addition to these 
indirect methods, there are others based on the direct determination
of this distance.  Measurements with trigonometric parallax of water
masers near the center of the galaxy might be the ideal method,
although with its current precision it is not
competitive~\cite{Reid:2009fu}.  Another method is based on
spectroscopic information of parallax and proper motions of regions of
high-mass star formation, along with modeling of the rotation
curve~\cite{Reid:2009nj,   McMillan:2009yr, Bovy:2009dr,
  Honma:2012zd}.  Recently, there has been some controversy regarding
the value of the solar motion component in the direction of the galactic
rotation~\cite{Schoenrich:2009bx}, which is correlated with $\Rsun$,
and on the sensitivity to the parametrization of the rotation curve.
This has been discussed in detail in a new analysis that has obtained
$\Rsun = 8. 34 \pm 0.16$~\cite{Reid:2014boa}.  Finally, astrometric
and radial velocity measurements of the Milky Way nuclear star
cluster, around the black hole in the galactic center, measure a
similar distance, $\Rsun \sim 8.3-8.5$~kpc~\cite{Ghez:2008ms,
  Gillessen:2008qv, Gillessen:2009ht, Do:2013}, although this
measurement is strongly correlated with the mass of the central black 
hole.  All in all, estimates based on direct distance measurements
seem to converge to $\Rsun \sim 8.3$~kpc, whereas more indirect
methods tend to obtain values in the range $\Rsun \sim 7.5-8.2$~kpc.
In this work, we consider the range $\Rsun=[7.5, \, 9.0]$~kpc, and use
a flat prior.

\section{Results}
\label{sec:results}

\begin{figure}[t]
  \centering
  \begin{center}
    \begin{tabular}{ll}
      \includegraphics[width=0.43\textwidth,angle=0]{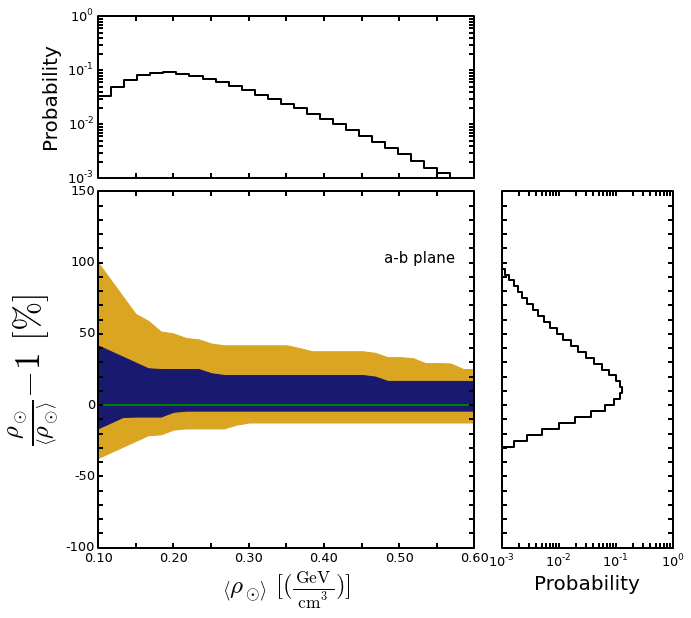} & 
      \includegraphics[width=0.43\textwidth,angle=0]{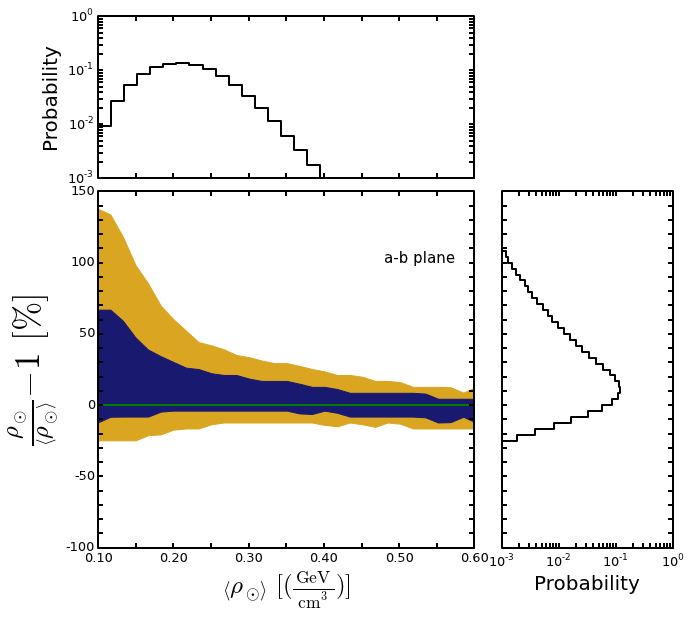}
      \\ 
      \includegraphics[width=0.43\textwidth,angle=0]{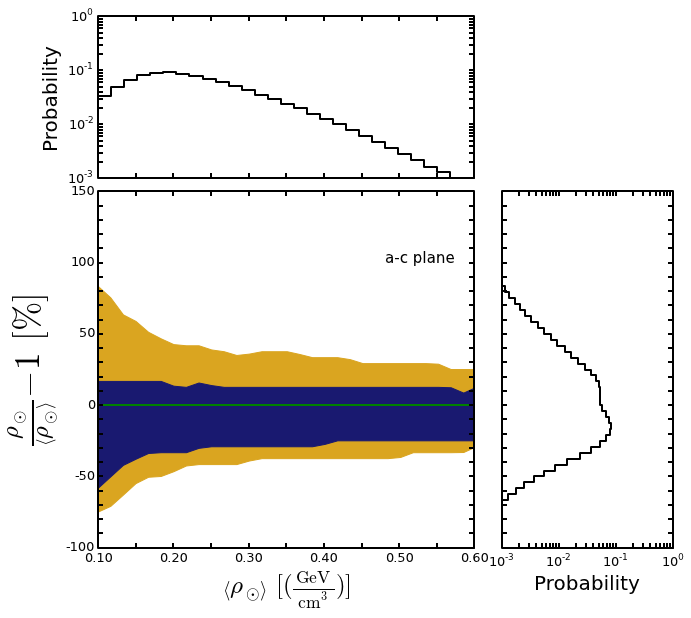} & 
      \includegraphics[width=0.43\textwidth,angle=0]{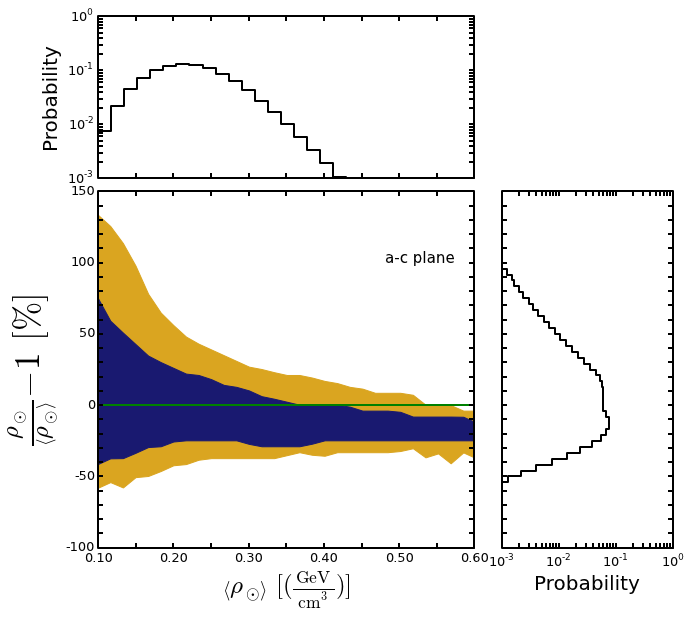}
      \\
      \includegraphics[width=0.43\textwidth,angle=0]{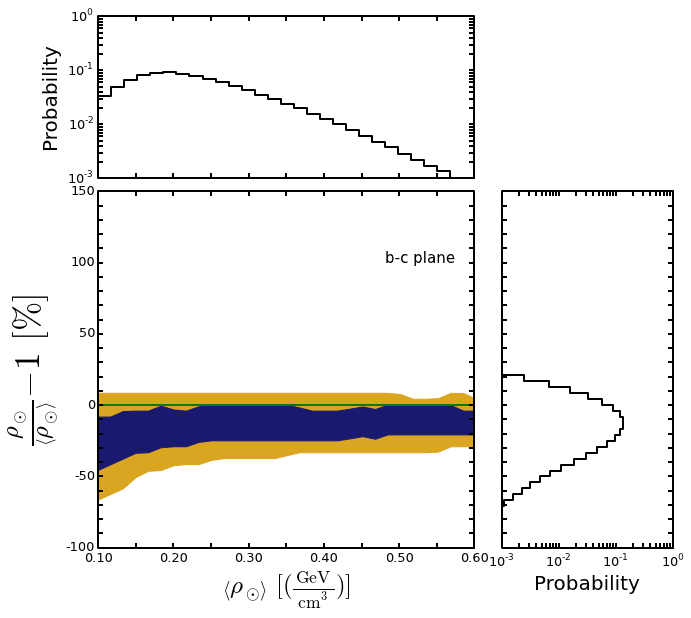} & 
      \includegraphics[width=0.43\textwidth,angle=0]{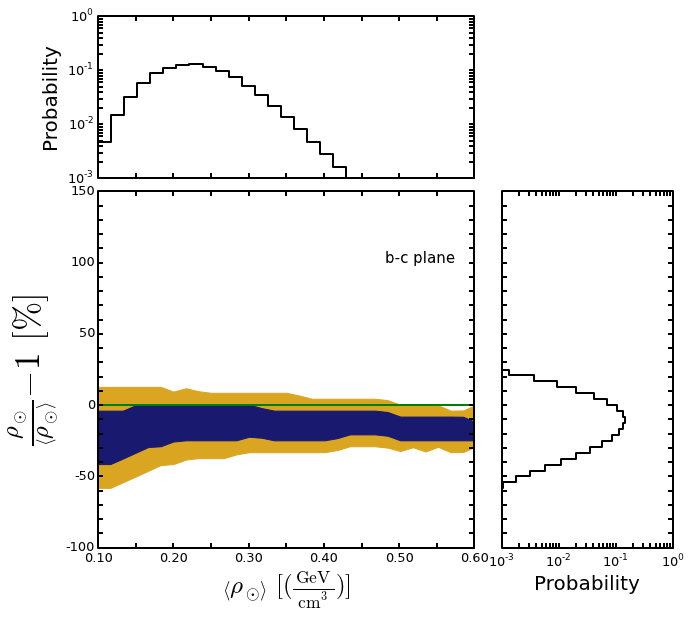}
      \\ 
    \end{tabular}
  \end{center}
  \caption{\textbf{\textit{Systematic uncertainties on
        $\boldsymbol{\rhosun}$, stemming from the non-sphericity of
        the Milky Way DM halo.}}  We show the probability distribution
    of the deviation of the local DM density from its spherically
    average value, $\langle \rhosun \rangle$, for the three symmetry
    planes as a function of $\langle \rhosun \rangle$.  The left
    (right) panels depict the results for the case without (with)
    priors included.  The dark blue (light orange) contours represent
    the $68\%$ ($95\%$) most probable regions.  On the top and the
    right of each panel we depict the projected probability
    distribution with respect to that quantity.}
  \label{fig:rho}
\end{figure}

Using the complete set of parameters, which define the ellipsoidal
NFW profile for each halo in the simulation sample, we evaluate the
probability distribution of variations with respect to the spherically
averaged quantities of interest in this work: the local DM density and
the so-called $J$ factors in indirect DM searches.  We show results
for the case when no priors are considered, i.e., just the bare
distributions from the simulation sample are used to quantify the
systematic effects.  And we also show results for the case when we make
use of observational information and analyze the data by adding priors
on different quantities, which weigh the original probability
distributions, as described above.

\subsection[Systematic uncertainties on $\rho_{\odot}$]{Systematic uncertainties on \boldmath{$\rho_{\odot}$}}  

As already mentioned, the value of $\rhosun$ deduced from most dynamic
measurements often refers to the spherically averaged density $\langle
\rhosun \rangle$.  We also noted that, in a non-spherical halo, the
actual DM density in the solar neighborhood could actually differ
significantly from that value (see Sec.~\ref{sec:impact}).  Here, we
consider the whole sample of halos from the {\it MultiDark Database}
and study this type of uncertainties as a function of the local average
of the DM density.

The spherically averaged local DM densities, $\langle \rhosun
\rangle$, at distance $\Rsun$ are computed for all halos in our data
set, using the parameters that fit each halo with the triaxial density
profile given by Eq.~(\ref{eq:nfwnsph}).  All halos are binned
according to their value of the spherical average $\langle \rhosun
\rangle$.  Then, for every halo, $\rhosun$ is evaluated in a grid of
300 different points along a circle of radius $\Rsun$ for each plane
of symmetry and for 6 values of $\Rsun$ covering the range in
Tab.~\ref{tab:priors}.  The results for the deviations from the 
spherically averaged value are depicted in Fig.~\ref{fig:rho}.
The left and right columns correspond to cases without and with priors, 
respectively.  In this figure, we show the deviation of
$\rhosun$ from its spherically averaged value $\langle \rhosun
\rangle$ for the three symmetry planes as a function of $\langle
\rhosun \rangle$.  The dark blue (light orange) contours represent the
$68\%$ ($95\%$) most probable regions of the deviation.  On the top
and the right of each panel the projected probability distributions of
$\langle\rhosun\rangle$ and $\frac{\rhosun}{\langle \rhosun \rangle}-1$
are depicted, respectively.

In the principal plane $a-b$, the local density tends to adopt values
larger than those of the average density, whereas in the plane $b-c$
the values are typically smaller than the average.  The deviations (in
absolute value) in the $a-b$ plane are larger compared to the ones in
the $b-c$ plane due to the fact that there are more prolate than
oblate halos (see Figs.~\ref{fig:triax} and~\ref{fig:haloimpact}).
The densities in the plane $a-c$ are intermediate, spanning over a
large range that goes from the lowest values reached in the $b-c$
plane to the highest values in the $a-b$ plane.

Fig.~\ref{fig:rho} shows important deviations with respect to the
spherically averaged local DM density, especially for low values,
$\langle \rhosun \rangle \lesssim 0.2$~GeV/cm$^3$.  Such small values
for $\langle \rhosun \rangle$ are common in very triaxial halos which,
in turn, naturally generate large deviations.  We also note that for
the case without priors, for values of the average local DM density
above $\sim 0.2$~GeV/cm$^3$ the deviations do not change significantly,
being of the order of $^{+20\%}_{-5\%}$
$\left(^{+40\%}_{-15\%}\right)$,  $^{+15\%}_{-30\%}$
$\left(^{+35\%}_{-40\%}\right)$ and $^{+0\%}_{-30\%}$
$\left(^{+10\%}_{-35\%}\right)$ for the $68\%$ ($95\%$) most probable
regions in the $a-b$ plane, $a-c$ plane and $b-c$ plane, respectively.  

A few differences are observed when priors are added. The
prior on $M_{60}$ mostly affects the probability distribution of
$\langle \rhosun \rangle$, which gets narrower and tends to peak
around $\sim 0.20$ - $0.25$ GeV/cm$^3$.  Let us recall that the prior
on $M_{60}$ is independent of the point of observation and only
depends on the global properties of the halo.  However, the
prior on the value of the surface density, $\Sigma_{1.1}^{\rm DM}$,
has a strong influence on the distribution of $\rhosun / \langle
\rhosun \rangle -1$.  The uncertainties on $\rhosun$ tend to be slightly
smaller for intermediate and high values of $\langle \rhosun \rangle$
and slightly larger for low values of $\langle \rhosun \rangle$,
although the differences, in general, are not very significant.

\begin{figure}[t]
  \centering
  \begin{center}
    \begin{tabular}{ll}
      \includegraphics[width=0.43\textwidth,angle=0]{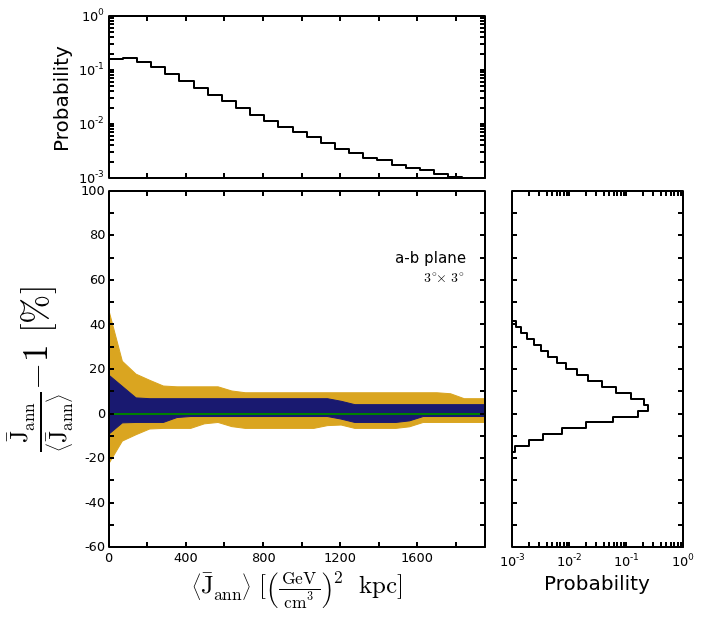} &  
      \includegraphics[width=0.43\textwidth,angle=0]{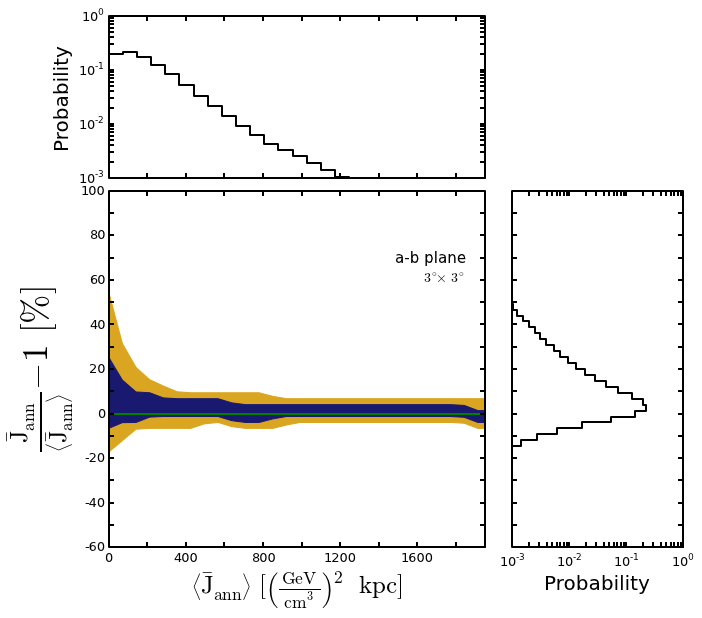}\\ 
      \includegraphics[width=0.43\textwidth,angle=0]{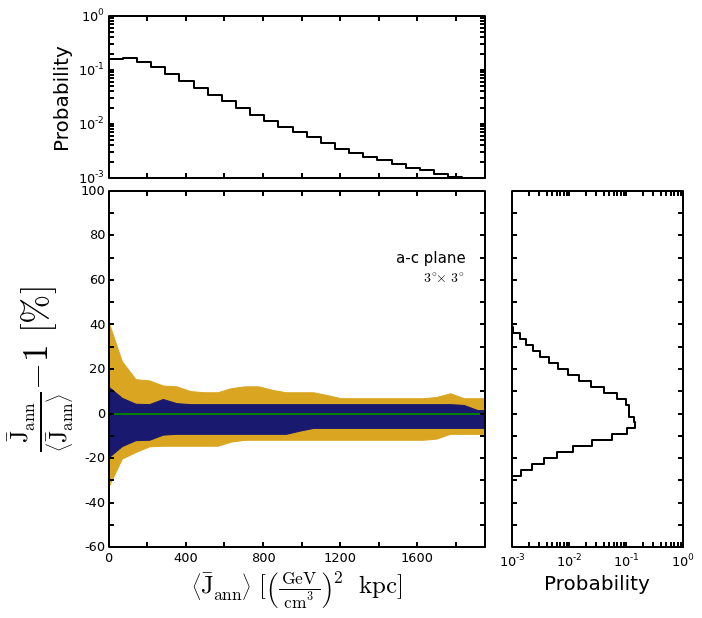} & 
      \includegraphics[width=0.43\textwidth,angle=0]{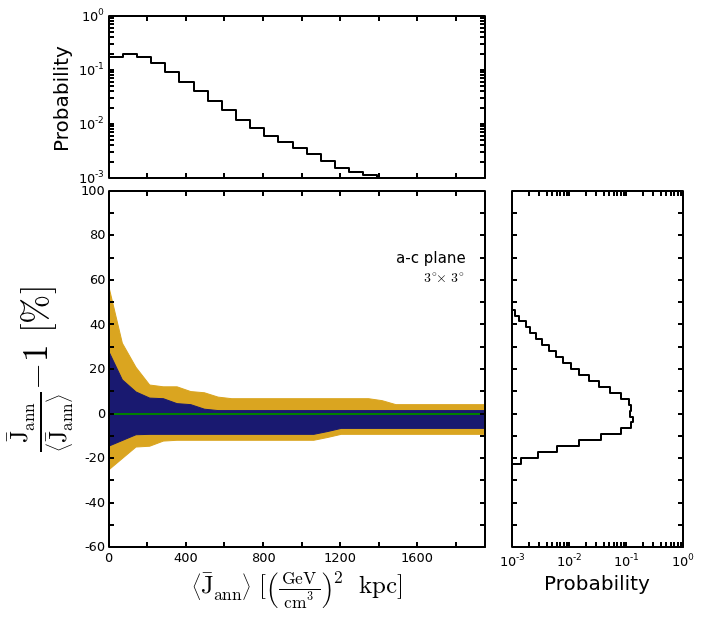}\\ 
      \includegraphics[width=0.43\textwidth,angle=0]{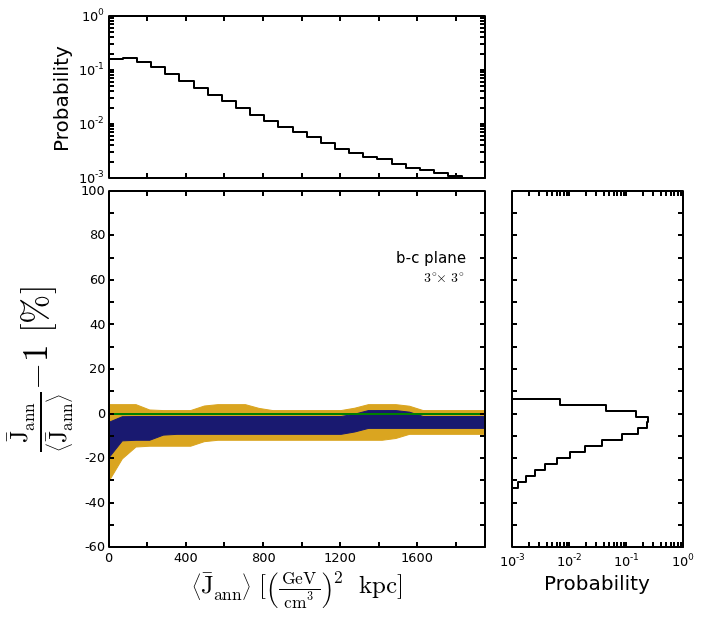} & 
      \includegraphics[width=0.43\textwidth,angle=0]{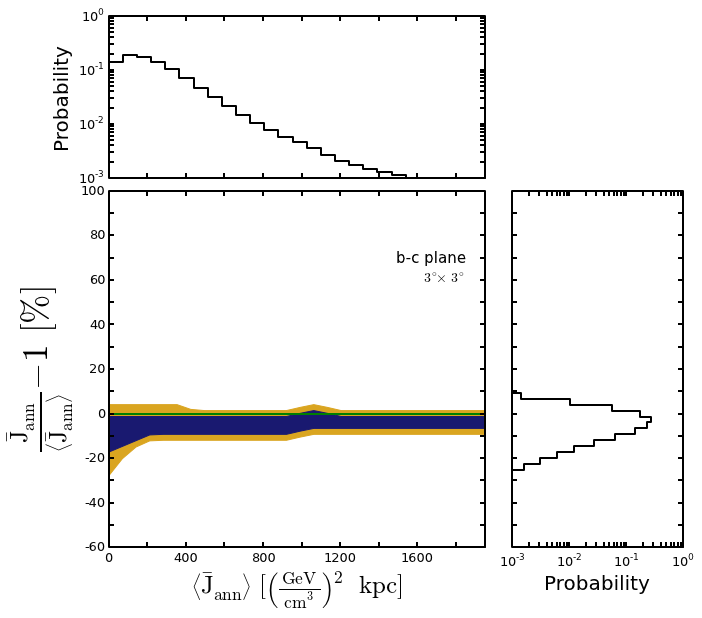}\\ 
    \end{tabular}
  \end{center}
  \caption{\textbf{\textit{Systematic uncertainties on
        $\boldsymbol{\bar{J}_{\rm ann}}$ for a square ROI of
        $\boldsymbol{3^\circ\times 3^\circ}$ around the galactic
        center for DM annihilations, stemming from the non-sphericity
        of the Milky Way DM halo.}}  We show the probability
    distribution of the deviation of $\bar{J}_{\rm ann}$ from its
    spherically average value, $\langle \bar{J}_{\rm ann} \rangle$,
    for the three symmetry planes as a function of $\langle
    \bar{J}_{\rm ann} \rangle$.  The panels and colors of the
    different contours represent the same as in
    Fig.~\ref{fig:rho}. \vspace{5mm}}
\label{fig:jann}
\end{figure}

\begin{figure}[t]
  \centering
  \begin{center}
    \begin{tabular}{ll}
      \includegraphics[width=0.43\textwidth,angle=0]{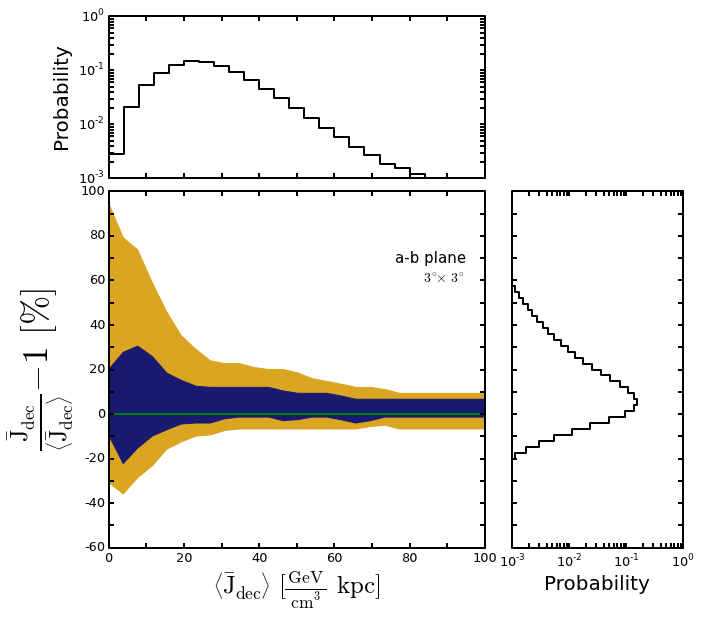} & 
      \includegraphics[width=0.43\textwidth,angle=0]{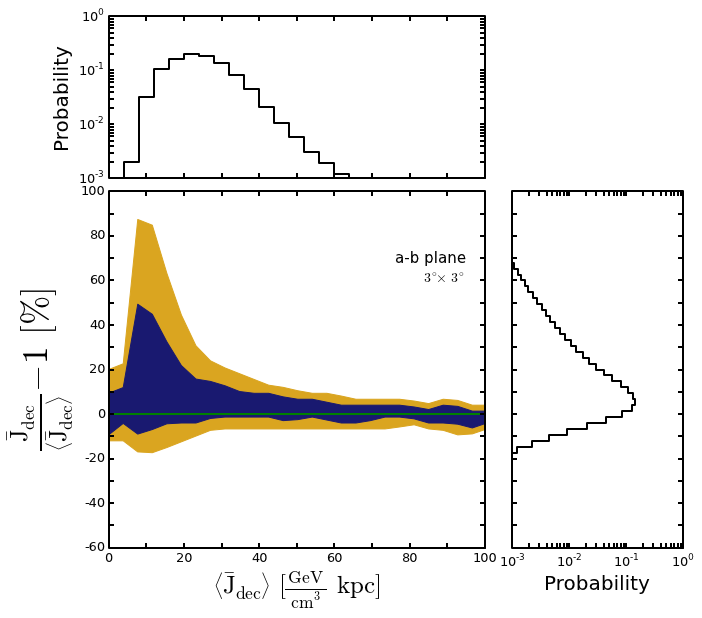}\\ 
      \includegraphics[width=0.43\textwidth,angle=0]{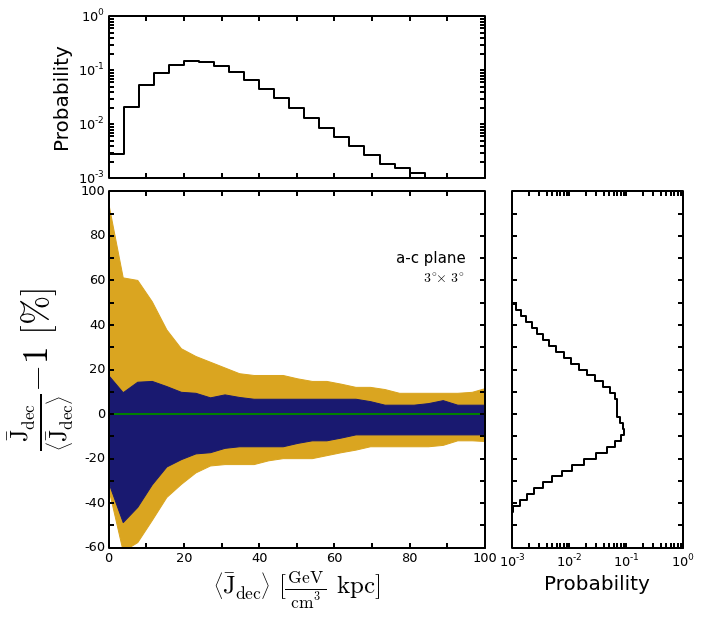} & 
      \includegraphics[width=0.43\textwidth,angle=0]{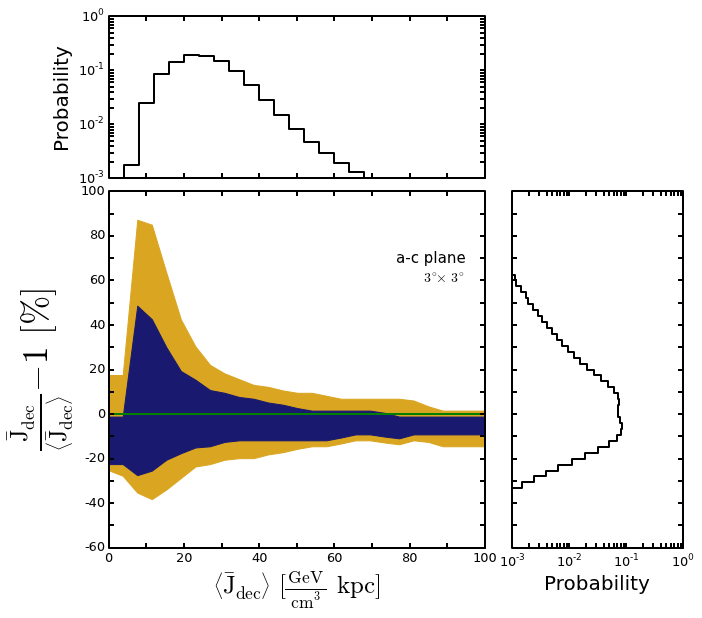}\\ 
      \includegraphics[width=0.43\textwidth,angle=0]{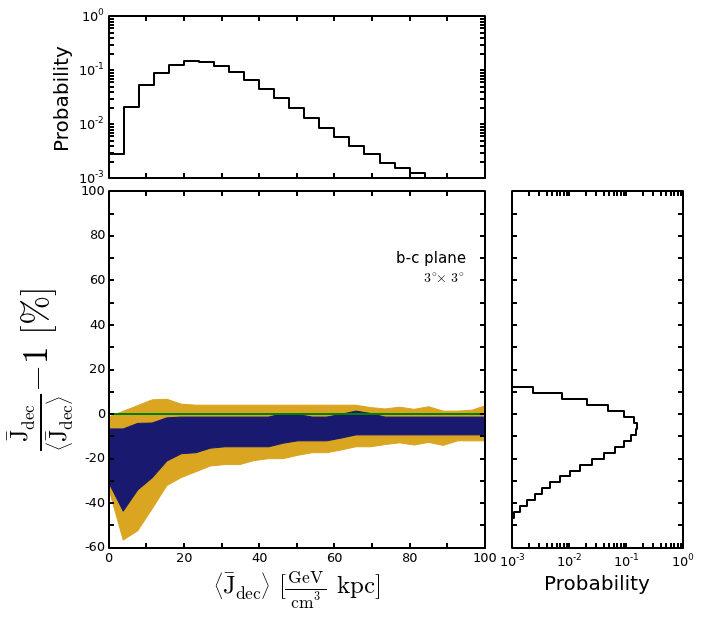} & 
      \includegraphics[width=0.43\textwidth,angle=0]{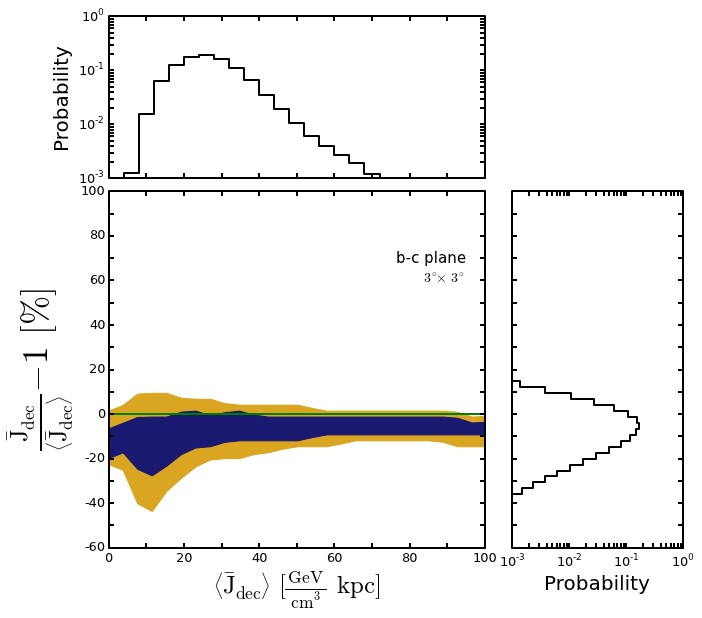}\\ 
    \end{tabular}
  \end{center}
  \caption{\textbf{\textit{Systematic uncertainties on
        $\boldsymbol{\bar{J}_{\rm dec}}$ for a square ROI of
        $\boldsymbol{3^\circ\times 3^\circ}$ around the galactic
        center for DM decays, stemming from the non-sphericity of the
        Milky Way DM halo.}}  We show the probability distribution of
    the deviation of $\bar{J}_{\rm dec}$ from its spherically average
    value, $\langle \bar{J}_{\rm dec} \rangle$, for the three symmetry
    planes as a function of $\langle \bar{J}_{\rm dec} \rangle$.  The
    panels and colors of the different contours represent the same as
    in Fig.~\ref{fig:rho}. \vspace{5mm}}
\label{fig:jdec}
\end{figure}

\subsection[Systematic uncertainties on $\bar{J}_{\rm ann}$ and
  $\bar{J}_{\rm dec}$]{Systematic uncertainties on
  \boldmath{$\bar{J}_{\rm ann}$} and \boldmath{$\bar{J}_{\rm dec}$}} 

Analogously to the analysis performed to estimate the uncertainties on
the local DM density (Fig.~\ref{fig:rho}), we also compute the
systematic uncertainties, caused by the non-sphericity of the Milky
Way DM halo, on the determination of the $J$ factors for DM
annihilation and decay.  Our results are depicted in
Fig.~\ref{fig:jann} for $\bar{J}_{\rm ann}$ and in Fig.~\ref{fig:jdec}
for $\bar{J}_{\rm dec}$, as a function of their spherically averages,
$\langle \bar{J}_{\rm ann} \rangle$ and $\langle \bar{J}_{\rm dec}
\rangle$, for the three symmetry planes and for a square ROI of
$3^\circ\times 3^\circ$ around the galactic center.  The left (right)
panels correspond to the case without (with) priors.  The blue
(orange) regions represent the $68\%$ ($95\%$) most probable contours.
These figures have essentially the same features of
Fig.~\ref{fig:rho}: for intermediate and high values of $\langle
\bar{J} \rangle$, approximately the same uncertainty is found, whereas
larger errors are obtained for low values\footnote{The bump in
  Fig.~\ref{fig:jdec}, which occurs at very low values of $\langle
  \bar{J}_{\rm dec} \rangle$, is due to the presence of a small number
  of approximately spherical halos with very large concentrations in
  the first bin, which further get favorably weighted by the priors.
  It is a statistical effect due to the finite size of the sample and 
  the chosen size of the bins.  It does not show up in
  Fig.~\ref{fig:jann} because the number of halos in the first bin is
  two orders of magnitude larger.} of $\langle \bar{J} \rangle$.  For
DM annihilations (Fig.~\ref{fig:jann}) and for intermediate and high
values of $\langle \bar{J}_{\rm ann} \rangle$, the deviations are of
the order of a few percent (up to $\sim$10\%) for the $68\%$ ($95\%$)
most probable regions.  On the other hand, for DM decays
(Fig.~\ref{fig:jdec}), the deviations from the average value for
intermediate and high $\langle \bar{J}_{\rm dec} \rangle$, are of the
order of a few percent, up to $\sim$10\% (up to $\sim$15\%) for the
$68\%$ ($95\%$) most probable regions.  The inclusion of priors has a 
very similar effect to what happens for the local DM density.  The
prior on $M_{60}^{\rm DM}$ gives more weight to low values of the $J$
factors and makes the probability distribution narrower, whereas the
prior on $\Sigma_{1.1}^{\rm DM}$ makes the variations of the $J$
factors to be slightly smaller for intermediate and high values and
larger for low values.  All in all, the inclusion of priors produces
small changes with respect to the results without priors.  Finally,
let us stress that we have verified that the variations with respect
to the average value for the $J$ factors depend very weakly on the
chosen ROI (around the galactic center).

In order to serve as a reference, let us note that the $J$ factors for
the same ROI, corresponding to a spherical NFW profile with
$r_s=20$~kpc and $\rhosun = 0.3$~GeV/cm$^3$ at $\Rsun = 8.3$~kpc, are:
$\langle \bar{J}_{\rm ann} \rangle = 590$~(GeV/cm$^3 )^2$~kpc and
$\langle \bar{J}_{\rm dec} \rangle= 43$~(GeV/cm$^3 )$~kpc.

\section{Discussion and Conclusions}
\label{sec:concl}

Direct and some indirect strategies of DM searches depend on its spatial
distribution in the galaxy.  If a DM signal is detected, one of the
main focus of these searches would be to deduce properties of the DM
particle.  Otherwise, the non-observation of a signal could be used to
derive upper limits on the DM annihilation and scattering cross
section.  In the case of DM direct detection, the event rate is
directly proportional to the local DM density, $\rhosun$, as described
in Sec.~\ref{sec:DDID}.  Therefore, any systematic error on $\rhosun$
translates into an error on the limits (or measurement) of the
scattering cross section.  Similarly, for indirect detection, the
gamma-ray and neutrino flux from DM annihilations or decays is
directly proportional to the so-called $J$ factors.  The error on them
directly translates into an error on the limits (or measurement) of the
thermally averaged DM annihilation cross section and on the DM mean
lifetime.

Although DM halo profiles are usually assumed to be spherical, it is
well known that N-body simulations predict halos to be
non-spherical~\cite{Frenk:1988zz, Katz:1991, Dubinski:1991bm,
  Warren:1992tr, Dubinski:1993df, Jing:2002np, Kasun:2004zb,
  Bailin:2004wu, Hopkins:2004np, Allgood:2005eu, Maccio':2006nu,
  Bett:2006zy, Hayashi:2006es, Kuhlen:2007ku, Stadel:2008pn,
  MunozCuartas:2010ig, VeraCiro:2011nb, Schneider:2011ed,
  Vera-Ciro:2014ita}.  In this work we consider a very large sample of
$\sim 10^5$ DM-only halos (described in Sec.~\ref{sec:simulations}),
from the N-body cosmological simulation
\verb"Bolshoi"~\cite{Klypin:2010qw}, publicly available through the
{\it MultiDark Database}, with masses in the range of that of the
Milky Way.  We construct the probability distributions of the
parameters that define their shape and use them to study the impact of
halo asphericity on the determination of the local DM density and the 
$J$ factors relevant for indirect searches of signals from the
galactic center.  This is first illustrated in Sec.~\ref{sec:impact}
with three example halos: an approximately spherical halo, a prolate
halo and an oblate halo.

In addition to the halo sample obtained from the N-body simulation, we
also add several observational constraints: on the virial mass, on the
enclosed mass at 60~kpc, on the local DM surface density and on the
Sun's galactocentric distance.  All these constraints and the way we
implement them are described in Sec.~\ref{sec:priors}.

Finally, in Sec.~\ref{sec:results} we show our results without and
with observational priors included, although the differences are
small.  We have shown that, including priors, for values of the
spherical average of the local DM density of the order of current
estimates, i.e., $\langle \rhosun \rangle \simeq 0.3-0.4$~GeV/cm$^3$,
the actual value of $\rhosun$, if the stellar disk coincides with the
$a-b$ plane of the DM halo, with a probability of 95\%, lies in the
interval (see upper right panel of Fig.~\ref{fig:rho})
\begin{equation}
  \frac{\rhosun}{\langle \rhosun \rangle} = 0.83-1.35 ~,
\end{equation}
in rough agreement with Refs.~\cite{Zemp:2008gw, Pato:2010yq}.  On the
other hand, if the stellar disk coincides with the $a-c$ ($b-c$) plane
the range, with a probability of 95\%, the range is $\rhosun/\langle
\rhosun \rangle = 0.62-1.27$ $(0.67-1.08)$.  Let us note, however,
that these two configurations are not stable~\cite{Debattista:2013rm}.

In a similar way, we have also computed the impact of halo asphericity
on the values of the $J$ factors relevant for indirect searches of
gamma-rays and neutrinos from DM annihilations and decays at the
galactic center.  We note that the variations with respect to the
average value depend very weakly on the chosen ROI (around the galactic
center).  However, in these cases the variation with respect to the
spherical averages are much smaller than those obtained for the local
density.  Including observational priors, for the case of DM
annihilations and $\langle \bar{J}_{\rm ann} \rangle \simeq
590$~(GeV/cm$^3 )^2$~kpc (for a square ROI of $3^\circ\times 3^\circ$
around the galactic center), the actual value of $\bar{J}_{\rm ann}$,
if the stellar disk coincides with the $a-b$ plane of the DM halo,
with a probability of 95\%, lies in the interval (see upper right
panel of Fig.~\ref{fig:jann})
\begin{equation}
  \frac{\bar{J}_{\rm ann}}{\langle \bar{J}_{\rm ann} \rangle} =
  0.95-1.09 ~,  
\end{equation}
whereas it is $\bar{J}_{\rm ann} / \langle \bar{J}_{\rm ann} \rangle =
0.90-1.07$ $(0.88-1.01)$, if the stellar disk coincides with the $a-c$
($b-c$) plane.

Including observational priors, for DM decays and $\langle
\bar{J}_{\rm dec} \rangle \simeq 43$~(GeV/cm$^3 )$~kpc (for the same
ROI), the actual value of $\bar{J}_{\rm dec}$, if the stellar disk
coincides with the $a-b$ plane of the DM halo, with a probability of
95\%, lies in the interval (see upper right panel of
Fig.~\ref{fig:jdec})
\begin{equation}
  \frac{\bar{J}_{\rm dec}}{\langle \bar{J}_{\rm dec} \rangle} =
  0.93-1.13 ~,
\end{equation}
whereas it is $\bar{J}_{\rm dec} / \langle \bar{J}_{\rm dec} \rangle
= 0.82-1.12$ $(0.83-1.04)$, if the stellar disk coincides with the
$a-c$ ($b-c$) plane. 

The ranges above are quoted for values of the spherical averages equal
to the $J$ factors for a spherical NFW profile with $r_s=20$~kpc and
$\rhosun = 0.3$~GeV/cm$^3$ at $\Rsun = 8.3$~kpc (for a square ROI of 
$3^\circ\times 3^\circ$ around the galactic center), i.e., $\langle
\bar{J}_{\rm ann} \rangle \simeq 590$~(GeV/cm$^3 )^2$~kpc and $\langle
\bar{J}_{\rm dec} \rangle \simeq 43$~(GeV/cm$^3 )$~kpc.  However, it
turns out that the corresponding values for other spherical DM
profiles span a larger range than that owing to halo asphericity.  For
instance, for a Einasto profile~\cite{Einasto:1965} with the same
local DM density and with $\alpha = 0.17$, $r_s = 20$~kpc, $\langle 
\bar{J}_{\rm ann} \rangle \simeq 10^3$~(GeV/cm$^3 )^2$~kpc and $\langle
\bar{J}_{\rm dec} \rangle \simeq 56$~(GeV/cm$^3 )$~kpc.  On the
other hand, for a Burkert profile~\cite{Burkert:1995yz} with the same
local DM density with $r_s = 12$~kpc, $\langle \bar{J}_{\rm ann}
\rangle \simeq 4.9$~(GeV/cm$^3 )^2$~kpc and $\langle \bar{J}_{\rm
  dec} \rangle \simeq 11$~(GeV/cm$^3 )$~kpc.  Let us note that the
less cuspy the DM profile the larger the variations of the $J$ factors
with respect to the spherical average~\cite{Athanassoula:2005dw},
which can be understood in the same way $\bar{J}_{\rm dec}$
experiences larger variations than $\bar{J}_{\rm ann}$ for a given
profile.  Thus, we conclude that uncertainties originated from
the non-sphericity of the Milky Way DM halo are smaller, and thus less
important, than the uncertainties coming from the DM density profile.

In summary, we have quantified the systematic uncertainties on the
local DM density and the $J$ factors in a statistical way by using the
results from the \verb"Bolshoi" simulation.  We note that halo
asphericity could imply systematic errors on the local DM density at
the level of current uncertainties, but in the case of the $J$ factors
they tend to be smaller than other errors.  The determination of the
DM density profile, not only is important for a better understanding
of our galaxy, but also because they represent crucial parameters in
direct and indirect DM searches.  Hence, assessing their systematic
uncertainties, and in particular due to halo asphericity, is of prime
importance.  Extracting DM properties from a positive signal or from a
combination of positive signals will critically depend on the value of
these parameters.

\acknowledgments
We would like to thank Manuel Drees for useful discussions and comments.
NB is supported by the S\~ao Paulo Research Foundation (FAPESP) under
grants 2011/11973-4 and 2013/01792-8.  JEF-R is supported by a FAPA
starting grant from the Vicerrector\'ia de Investigaciones at
Universidad de los Andes in Bogot\'a, Colombia.  RG is supported by
the DFG TRR33 `The Dark Universe' and by the Helmholtz Alliance for
Astroparticle Physics.  SPR is supported by a Ram\'on y Cajal contract
and by the Spanish MINECO under grant FPA2011-23596.  SPR is also
partially supported by PITN-GA-2011-289442-INVISIBLES and the
Portuguese FCT through the projects CERN/FP/123580/2011,
PTDC/FIS-NUC/0548/2012 and CFTP-FCT Unit 777
(PEst-OE/FIS/UI0777/2013), which are partially funded through POCTI 
(FEDER).

\small
\bibliographystyle{JHEP}
\bibliography{biblio}

\end{document}